   \newcommand{\be}[0]{\begin{equation}}
   \newcommand{\ee}[0]{\end{equation}}
   \newcommand{\ba}[0]{\begin{eqnarray}}
   \newcommand{\ea}[0]{\end{eqnarray}}  

	\newcommand{\MS}[0]{\overline{{\rm MS}} }
	
	\newcommand{\AC}[0]{analytical continuation }
	\newcommand{\EC}[0]{effective charge }
\documentstyle[12pt,epsfig]{article}

\begin{document}

\topmargin=-0.4in
\headheight=0.6in        
\headsep=0in
\footheight=0.5in
\footskip=0.5in

\begin{titlepage}

\Large
\hfill\vbox{\hbox{DTP/97/30}
            \hbox{May 1997}}
\nopagebreak

\vspace{0.75cm}
\begin{center}
\LARGE
{\bf The uncertainty in $\alpha_{s}(M_{Z}^{2})$
determined from hadronic tau decay measurements}
\vspace{0.6cm}
\Large

C.J. Maxwell and D.G. Tonge

\vspace{0.4cm}
\large
\begin{em}
Centre for Particle Theory, University of Durham\\
South Road, Durham, DH1 3LE, UK
\end{em}

\vspace{1.7cm}

\end{center}
\normalsize
\vspace{0.45cm}

\begin{abstract}

We show that QCD Minkowski observables such as the $e^{+}e^{-}$
$R$-ratio and the hadronic tau decay $R_{\tau}$ are completely
determined by the effective charge (EC) beta-function, $\rho(x)$,
corresponding to the Euclidean QCD vacuum polarization
Adler $D$-function, together with the next-to-leading order
(NLO) perturbative coefficient of $D$. An efficient numerical
algorithm is given for evaluating $R$, $R_{\tau}$ from a 
weighted contour integration of $D(se^{i\theta})$ around
a circle in the complex squared energy $s$-plane,
with $\rho(x)$ used to evolve in $s$ around the contour.
The EC beta-function can be truncated at next-to-NLO (NNLO)
using the known exact perturbative calculation or the 
uncalculated $\rm{N}^{3}$LO and higher terms can be
approximated by the portion containing the highest
power of $b$, the first QCD beta-function coefficient.
The difference between the $R$, $R_{\tau}$
constructed using the NNLO and ``leading-b" resummed
versions of $\rho(x)$ provides an estimate of the uncertainty
due to the uncalculated higher order corrections. Simple
numerical parametrizations are given to facilitate these
fits. For $R_{\tau}$ we estimate an uncertainty 
$\delta\alpha_{s}(m_{\tau}^{2})\simeq0.01$, corresponding
to $\delta\alpha_{s}(M_{Z}^{2})\simeq0.002$. This encouragingly
small uncertainty is much less than rather pessimistic 
estimates by other authors based on analogous all-orders
resummations, which we demonstrate to be extremely dependent
on the chosen renormalization scheme, and hence misleading.

\end{abstract}



\end{titlepage}

\newpage

\section{Introduction}

There has been extensive recent interest [1--3] in the possibility
of using measurements of $R_{\tau}$, the total hadronic decay width
of the $\tau$ lepton normalized to the leptonic decay width,
for precise determination of the renormalized strong coupling
$\alpha_{s}(M_{Z}^{2})$ (or more fundamentally $\Lambda_{\MS}$). For
this purpose $R_{\tau}$ apparently possesses a number of advantages
compared with other QCD observables. It is an inclusive quantity
which can be computed in QCD using the operator product expansion 
(OPE) supplemented by analyticity [4--7]. 
It has been calculated in QCD perturbation theory to 
next-to-next-to-leading order (NNLO) $O(\alpha_{s}^{3})$
\cite {kat,gorish1}. Power corrections are expected to be small 
\cite {braaten1,braaten2}, and since the $\tau$ mass is below
the threshold for charmed hadron production only the
light quarks $u$, $d$, $s$ are active, so QCD with $N_{f}=3$ massless
quark
flavours should be applicable. $R_{\tau}$ can be rather accurately 
determined
from the measured electronic branching ratio of the $\tau$ or from 
the $\tau$ lifetime \cite {web}. In evolving up in energy scale from 
$\alpha_{s}(m_{\tau}^{2})$ to $\alpha_{s}(M_{Z}^{2})$, which is 
customarily
quoted in global comparisons, the fractional error in 
$\alpha_{s}(\Lambda_{\MS})$ is reduced. Measurement of the hadronic
width of the $Z^{0}$ to directly determine $\alpha_{s}(M_{z}^{2})$
shares the same advantages of being inclusive, calculated to NNLO
in perturbation theory and having small power corrections, but 
suffers
significant corrections from heavy quark masses, and much larger 
systematic
experimental errors.

Despite these undoubted advantages possessed by $R_{\tau}$ as a means
of determining $\alpha_{s}$, the relatively low energy scale
involved, $s=m_{\tau}^{2}$, might lead one to expect sizeable 
corrections
from uncalculated $O(\alpha_{s}^{4})$ and higher orders in 
perturbation
theory. To assess the effect of these terms with our present limited
state of knowledge one can employ a, necessarily approximate, 
all-orders
resummation of the QCD perturbation series. A well-motivated framework
for this is provided by the leading-$b$ approximation 
\cite {charles1,charles2}, also 
sometimes referred to as naive non-abelianization 
\cite {broad2,benbraun1}. In this approach
the portion of perturbative coeffcients containing the highest power
of $b=\frac{1}{6}(11N$--$2N_{f})$, the first beta-function coefficient
for SU($N$) QCD with $N_{f}$ active quark flavours, is resummmed to 
all-orders. This can be accomplished given exact large-$N_{f}$ 
all-orders results [15--17]. This technique has been applied to the 
QCD vacuum polarization Adler $D$-function \cite {broad1,ben1}, 
and its
Minkowski continuations, the $e^{+}e^{-}$ $R$-ratio and $R_{\tau}$.
Deep inelastic scattering (DIS) sum rules \cite {broadkat} and 
heavy quark
decay widths and pole masses
have also been discussed  \cite {ball,bert4}. 

In several recent papers it has been claimed that applying the 
leading-$b$
resummation to $R_{\tau}$ indicates rather large perturbative 
uncertainties \cite {ball,bert2,bert3}.
Indeed the estimated uncertainty in $\alpha_{s}(M_{Z}^{2})$ is of the
same order as that normally quoted in determinations
from jet observables at LEP and SLD \cite {burrows}.

In a recent paper \cite {tonge1} we have pointed out that a 
straightforward 
resummation of the leading-$b$ terms of the kind employed in 
references
\cite {ball,bert2,bert3}, is renormalization scheme (RS) dependent.
This occurs because
the compensation mechanism between the renormalization group (RG) 
improved coupling and the perturbative coefficients is destroyed
by retaining only the leading-$b$ terms. As a result the `naive' 
leading-$b$ resummation is not RS-invariant under the full QCD
renormalization group (RG).
Whilst at large energies the resulting ambiguities are mild, 
at $s=m_{\tau}^{2}$ this RS dependence is serious and in our view
invalidates the rather pessimistic conclusions of these references
regarding the likely uncertainty in $\alpha_{s}(M_{Z}^{2})$ determined
from $R_{\tau}$.

In reference \cite {tonge1} we proposed an improved RS-invariant 
resummation
based on approximating the unknown effective charge  (EC) 
beta-function
coefficients by the portion containing the highest power
of $b$. Approximated perturbative coefficients in any RS can then be
obtained using the exact QCD RG. The leading-$b$ effective charge
beta-function can be resummed using exact all-orders large-$N_{f}$
results.

The difference between the exact NNLO result for $R_{\tau}$ in the
effective charge RS, and the RS-invariant all-orders resummation
indicates a rather small uncertainty due to the approximated
higher order terms, and the estimated uncertainty in 
$\alpha_{s}(M_{Z}^{2})$
is correspondingly small, $\delta\alpha_{s}(M_{Z}^{2})\simeq0.001$.

In this paper we wish to explore the perturbative uncertainty in
$R_{\tau}$ in somewhat more detail. Both $R_{\tau}$ and
$e^{+}e^{-}$ $R$-ratio can be represented by a contour integral
involving $D(se^{i\theta})$, where $D(-s)$ is the Euclidean
Adler $D$-function, around a circle, cut along the positive
real axis,
$\theta=-\pi$ to 
$\theta=\pi$, in the complex $s$-plane \cite {schilder}.
Here $s=m_{\tau}^{2}$
for $R_{\tau}$. Conventional perturbation theory involves an expansion
in $\alpha_{s}(s)$ obtained by re-expressing 
$\alpha_{s}(se^{i\theta})$ as an
expansion in $\alpha_{s}(s)$ which is then truncated. Alternatively
one can simply numerically perform the contour integration for the
$\alpha_{s}^{k}(se^{i\theta})$ terms up to a given order \cite {lp}.
This
procedure includes in addition to conventional fixed-order
perturbation theory a resummation of analytical continuation terms.
A subset of these terms involve powers of the first beta-function
coefficient, $b$, together with $\pi^2$ factors, and are resummmed
to all-orders in the leading-$b$ approach. In addition, however,
there are potentially large contributions involving higher 
beta-function coefficients \cite {kat2}. It would seem sensible, 
therefore
to perform the improved RS-invariant resummation for 
$D(se^{i\theta})$,
and numerically evaluate the contour integral. In this way additional
analytical continuation terms not captured in the leading-$b$
resummation are included exactly. This can then be compared with the 
exact NNLO result for $D(se^{i\theta})$ in the EC scheme,
with the contour integral again numerically evaluated. Since in both
cases the \AC terms are resummed the difference should be indicative
of the effect
of the approximated higher \EC RS invariants for $D$ beyond NNLO.

The plan of the paper is as follows. In section 2 we shall intoduce 
the contour integral representations of the $e^{+}e^{-}$ $R$-ratio 
and $R_{\tau}$ in terms of the Adler $D$-function. Using Taylor's
theorem we can then expand $R$ and $R_{\tau}$ in terms of $D(s)$
and its energy derivatives, which in turn can be expressed in terms
of the effective charge beta-function for $D$ and its derivatives.
These results can be easily used to express the perturbative 
coefficients of the Minkowski quantities $R$ and $R_{\tau}$
in terms of those of the Euclidean Adler $D$-function and its
\EC RS invariants. We have compared these with existing expressions
available in the literature \cite {kat2}. We also derive relations 
between
the EC invariants for $R$ and $R_{\tau}$ and these for $D$. 
In section
3 we briefly review the basis of the RS-invariant resummation 
proposal
\cite {tonge1}. The contour integrals for the $R$ and $R_{\tau}$ 
are evaluated
by using Taylor's theorem succesively to evaluate $D$ at a series of 
values of complex $s$ around the unit circle contour of  integration.
A Simpson's rule numerical integration is then performed. 
The translation
of $D$  in complex $s$ involves the effective charge beta-function
and its derivatives. This function can be truncated or its 
leading-$b$
terms resummed \cite {tonge1}. In section 4 fits to the 
experimental data
for $\tilde R$ and $\tilde R_{\tau}$ are performed to determine
$\alpha_{s}$ from fixed-order and resummed perturbation theory.
In section 5 we conclude by
comparing the resulting values and estimates of the perturbative
uncertainty with those suggested by other approaches.

\section{Contour integral representation of Minkowski observables}

The two quantities with which we shall be concerned are defined as
follows.

The $e^{+}e^{-}$ $R$-ratio is the observable
\be
R\equiv\frac{\sigma(\mbox{e}^{+}\mbox{e}^{-}\rightarrow 
\mbox{hadrons})}
{\sigma(\mbox{e}^{+}\mbox{e}^{-}\rightarrow \mu^{+}\mu^{-})}\;.
\ee
In SU($N$) QCD perturbation theory
\be
R(s)=N\sum_{f}Q_{f}^{2}\left(1+\frac{3}{4}C_{F}\tilde{R}\right)
+\left(\sum_{f}Q_{f}
\right)^{2}\tilde{\tilde{R}}\;,
\ee
where $Q_{f}$ denotes the electric charge of the quarks and the 
summation is over the flavours accessible at a given energy. 
$s$ is the physical timelike Minkowski squared momentum transfer.
The SU($N$) Casimirs are defined as $C_{A}=N$, 
$C_{F}=(N^{2}-1)/2N$.

$\tilde{R}$ denotes the perturbative corrections to the parton model
result and has the formal expansion
\be
\tilde{R}(s)=a+r_{1}a^{2}+r_{2}a^{3}+\cdots+
r_{k}a^{k+1}+\cdots\;,
\ee
where $a\equiv\alpha_{s}(\mu^{2})/\pi$ denotes the renormalization 
group
(RG) improved coupling. The $\MS$ scheme with $\mu^{2}=s$ is often 
chosen.
The $\tilde{\tilde{R}}$ contribution first enters at 
O($a^{3}$) due to the existence of diagrams of the ``light-by-light'' 
type. 

The ratio $R_{\tau}$ is defined analogously using the total $\tau$ 
hadronic width as
\be
R_{\tau}\equiv\frac{\Gamma(\tau\rightarrow \nu_{\tau}+\mbox{hadrons})}
{\Gamma(\tau\rightarrow \nu_{\tau}e^{-}\overline{\nu}_{e})}\;.
\ee
Its perturbative expansion has the form
\be
R_{\tau}=N(|V_{ud}|^{2}+|V_{us}|^{2})\,S_{EW}\left[1+\frac{5}{12}
\frac{\alpha(m_{\tau}^{2})}{\pi} +\tilde{R}_{\tau}\right]
\;,
\ee
where the $V_{ud}$ and $V_{us}$ are CKM mixing
matrix elements, with $|V_{ud}|^{2}+|V_{us}|^{2}\approx1$.
Since the energy scale, $s=m_{\tau}^{2}$, of the process lies 
below the threshold for charmed hadron production only
three flavours $u$, $d$, $s$, are active. $\alpha(m_{\tau}^{2})$
denotes the electromagnetic coupling \cite {braaten3} and 
$S_{EW}\simeq1.0194$ \cite {sirlin} denotes further electroweak 
corrections.
$\tilde{R}_{\tau}$ has a perturbative expansion of the form of 
equation (3)
with coefficients which we shall denote $r_{k}^{\tau}$. There
are no ``light-by-light"corrections for $R_{\tau}$
since $(\Sigma Q_{f})^{2}=0$ for $u$, $d$, $s$ active
quark flavours.

These two Minkowski quantities can both be expressed in terms of 
the transverse
part of the correlator $\Pi(s)$ of two vector currents in the 
Euclidean region,
\be
\Pi(s)(q_{\mu}q_{\nu}-g_{\mu\nu}q^{2})=4\pi^{2}i
\int d^{4}x\,{\rm e}^{iq\cdot x}\langle 0|T\{j_{\mu}(x)
j_{\nu}(0)\}|0\rangle
\;,
\ee
where $s=-q^{2}>0$. 
In order to avoid an unspecified constant, it is convenient to
consider the related Adler $D$-function,
\be
D(s)=-s\frac{d}{ds}\Pi(s)\;.
\ee
In perturbation theory $D$ can be written in the form of equation (2)
involving perturbative corrections $\tilde{D}$ with an expansion
as in equation (3) involving coefficients $d_{k}$, and 
``light-by-light'' corrections $\tilde{\tilde{D}}$.

$\tilde{R}$, $\tilde{R}_{\tau}$ and related Minkowski quantities such 
as spectral moments \cite {lp} can all be written in terms of a 
weighted contour integral
of $\tilde{D}(s\mbox{e}^{i\theta})$ around a unit circle in the 
complex $s$-plane \cite {schilder}.

Denoting such a generic Minkowski observable as $\hat{R}$ we have
\be
\hat{R}(s_{0})=\frac{1}{2\pi}\int_{-\pi}^{\pi}
\mbox{d}\theta\,W(\theta)\,\tilde{D}(s_{0}\mbox{e}^{i\theta})\;
\ee
where $W(\theta)$ is a weight function which depends on the observable
$\hat{R}$. For $\tilde{R}$ we have $W(\theta)=1$, and for 
$\tilde{R}_{\tau}$
\be
W_{\tau}(\theta)=(1+2\mbox{e}^{i\theta}-2\mbox{e}^{3i\theta}-
\mbox{e}^{4i\theta})\;,
\ee
and $s_{0}=m_{\tau}^{2}$.

A novel representation for $\hat{R}$ can be obtained by using Taylor's
theorem to expand $\tilde{D}(s\mbox{e}^{i\theta})$ around $s=s_{0}$.
This yields 
\be
\hat{R}(s_{0})=\frac{1}{2\pi}\int_{-\pi}^{\pi}
\mbox{d}\theta\,W(\theta)\,\Bigg{\{}\tilde{D}(s_{0})+
\sum_{n=1}^{\infty}
i^{n}\frac{\theta^{n}}{n!} \frac{\mbox{d}^{n}}{{\mbox{d}\ln{s}}^{n}}
\tilde{D}(s)\Bigg|_{s=s_{0}}\Bigg{\}}\;.
\ee
The derivatives in equation (10) can be recast in terms of the 
effective 
charge beta-function $\rho(\tilde{D})$ \cite {grun1,reader}, 
and its derivatives. 
$\rho(\tilde{D})$ is defined by
\ba
\frac{\mbox{d}\tilde{D}(s)}{\mbox{d}\ln{s}}&=&-\frac{b}{2}\,
\rho(\tilde{D})\;\nonumber\\
&\equiv&-\frac{b}{2}(\tilde{D}^{2}+c\tilde{D}^{3}+
\rho_{2}\tilde{D}^{4}+\cdots
+\rho_{k}\tilde{D}^{k+2}+\cdots)\;.
\ea
Here $b=\frac{1}{6}(11N$--$2N_{f})$ is the first coefficient of the 
beta-function, and
\be 
c=\left[-\frac{7}{8}\frac{C_{A}^{2}}{b}-\frac{11}{8}
\frac{C_{A}C_{F}}{b}
+\frac{5}{4}C_{A}+\frac{3}{4}C_{F}\right]\;,
\ee
is the second universal beta-function coefficient. 
The higher coefficients
$\rho_{2}$, $\rho_{3}$, $\cdots$, in equation (11) are RS-invariants
and may be expressed in terms of the perturbative coefficients 
of $\tilde{D}$,
$d_{k}$, together with the beta-function coefficients, 
$c_{k}$, which define
the renormalization scheme employed to define the RG improved coupling
$a$ \cite {stev1}. Thus
\be
\frac{\mbox{d}a(\mu^{2})}{\mbox{d}\ln{\mu}^2}=-\frac{b}{2}\,(a^{2}+
ca^{3}+c_{2}a^{4}+\cdots+c_{k}a^{k+2}+\cdots)\;.
\ee
The \EC (EC) scheme corresponds to the choice of coupling 
$\tilde{D}=a$.
The first two EC invariants are given by 
\begin{eqnarray}
\rho_{2}&=&c_{2}+d_{2}-cd_{1}-d_{1}^2 \nonumber\\
\rho_{3}&=&c_{3}+2d_{3}-4d_{1}d_{2}-2d_{1}\rho_{2}-
cd_{1}^{2}+2d_{1}^{3}\;.
\end{eqnarray}
We note that in references \cite {tonge1,reader}, to which the 
reader is referred for
additional discussion of the EC beta-function, the dependent energy
variable was taken to be $Q$, whereas we are employing $s=Q^{2}$
in this paper, hence there are additional factors of $\frac{1}{2}$
in equations (11) and (13).

Using equation (11) one can then show that the energy derivatives
in equation (10) can be rewritten as
\be
\frac{\mbox{d}^{n}}{\mbox{d}{\ln{s}}^n}\tilde{D}(s)\Bigg|_{s=s_{0}}=
\Bigg{(}-\frac{b}{2}\Bigg{)}^{n}\Bigg{[}\rho(x)\frac{\mbox{d}}
{\mbox{d}x}\Bigg{]}^{n-1}\hspace{-6mm}\rho(x)
\Bigg|_{x=\tilde{D}(s_{0})},\,n>0\;.
\ee
Thus finally we can write equation (10) in the form
\be
\hat{R}(s_{0})=\tilde{D}(s_{0})+\sum_{n=1}^{\infty}
\Bigg{(}\frac{-ib}{2}\Bigg{)}^{n}
\frac{w_{n}}{n!}\Bigg{[}\rho(x)\frac{\mbox{d}}{\mbox{d}x}
\Bigg{]}^{n-1}\hspace{-6mm}
\rho(x)
\Bigg|_{x=\tilde{D}(s_{0})}\;.
\ee
Here $w_{n}$ denotes moments of the weight function $W(\theta)$,
\be
w_{n}=\frac{1}{2\pi}\int_{-\pi}^{\pi}\mbox{d}\theta\,\theta^{n}\,
W(\theta)\;.
\ee
For $\tilde{R}$ setting $W(\theta)=1$ yields $w_{n}=\pi^{n}/(n+1)$
($n$ even), $w_{n}=0$ ($n$ odd). The first two terms in the sum of
equation (16) are then
\be
\tilde{R}(s_{0})=\tilde{D}(s_{0})-\frac{\pi^{2}b^{2}}{24}\rho
\rho^{\prime}+
\frac{\pi^{4}b^{4}}{1920}\rho(\rho^{\prime3}+4\rho\rho^{\prime}
\rho^{\prime\prime}+\rho^{2}\rho^{\prime\prime\prime})+\cdots\;.
\ee
Primes denote differentiation of $\rho(x)$ with respect to $x$, 
evaluated
at $x=\tilde{D}(s_{0})$. Successive terms are RS-invariants resulting
from the resummation to all-orders of analytical continuation terms
proportional to $\pi^{2}b^{2}$, $\pi^{4}b^{4}$, $\cdots$, 
respectively.

For $\tilde{R}_{\tau}$ the weight function $W_{\tau}(\theta)$ of 
equation (9) has moments $w_{1}^{\tau}=\frac{19i}{12}$, 
$w_{2}^{\tau}=\frac{\pi^{2}}{3}-\frac{265}{72}$, $\cdots$. Then 
equation (16) yields
\be
\tilde{R}_{\tau}=\tilde{D}(m_{\tau}^{2})+\frac{19b}{24}\rho-
\Bigg{(}\frac{\pi^{2}}{3}-\frac{265}{72}\Bigg{)}\frac{b^{2}}{8}
\rho\rho^{\prime}+\cdots\;.
\ee 
>From equation (16) we see that Minkowski observables are
naturally expressed in terms of the Euclidean Adler $D$-function
and its EC beta-function. Given an all-orders definition of $\rho(x)$
one can discuss the radius of convergence of the sum in equation (16)
in $\tilde{D}(s_{0})$. This is an interesting question which
could be directly addressed using the leading-$b$ resummmation of
$\rho$ on which the RS-invariant resummations of reference 
\cite {tonge1}
are based. However, one can anticipate a rather restricted radius
of convergence by making the one-loop approximation $\rho(x)=x^{2}$.
One then finds
\be
\tilde{R}_{\mbox{\tiny{one-loop}}}=\frac{2}{b\pi}\mbox{arctan}
(\frac{b\pi\tilde{D}}{2})\;,
\ee
for the result of resumming the analytical continuation terms which
only involve the lowest beta-function coefficient. This suggests
that the radius of convergence is limited by 
$\tilde{D}<\frac{2}{b\pi}$
\cite {lp,piv2}. For $N_{f}=3$ this gives a radius of convergence 
$\frac{4}{9\pi}\simeq0.141\cdots$, which is to be compared with
$\tilde{D}(m_{\tau}^{2})\simeq0.1$ So that the expansion will not
be useful for evaluating $\tilde{R}_{\tau}$ using the leading-$b$
resummation of $\rho(x)$. As we shall discuss in the next section,
however, we shall be able to make use of the Taylor's theorem 
approach to evaluate $\tilde{D}(se^{i\theta})$ at closely spaced
intervals around the integration contour using the resummed $\rho(x)$.

To conclude this section we note that the expansion of equation (16)
is of use in straightforwardly relating the $\tilde{R}$, 
$\tilde{R}_{\tau}$
Minkowski perturbative coefficients $r_{k}$, $r_{k}^{\tau}$ 
to the Euclidean
$d_{k}$ coefficients of $\tilde{D}$. One simply expands equation (16)
as a power series in $\tilde{D}$ then on substituting
$\tilde{D}=a+d_{1}a^{2}+d_{2}a^{3}+\cdots$, and isolating 
the coefficient
of $a^{k+1}$ one can directly obtain $r_{k}$, $r_{k}^{\tau}$ in terms
of $d_{i\leq k}$, $\rho_{i\leq k}$ and $c$. The resulting calculated 
expressions are in agreement with the results available
in the literature for $k\leq5$ \cite {kat2}, on  using 
equations (14) to 
re-express beta-function coefficients $c_{i}$ in terms of $\rho_{i}$
invariants. 

In clarifying the connection between the various versions of 
fixed-order
perturbation theory to be compared in section 4 it will be useful
to relate the EC RS-invariants $\rho_{k}^{R}$ and 
$\rho_{k}^{R_{\tau}}$
corresponding to $\tilde{R}$ and $\tilde{R}_{\tau}$ to the 
$\rho_{k}^{D}$
(hitherto $\rho_{k}$) connected with $\tilde{D}$ \cite {racz}. 
This can easily be
accomplished by first evaluating $r_{k}$ and $r_{k}^{\tau}$ in the
EC scheme for $\tilde{D}$, so that $d_{1}=d_{2}=\cdots=0$.
These $r_{k}(d_{i}=0)$ and $r_{k}^{\tau}(d_{i}=0)$ are simply the
coefficient of $\tilde{D}^{k+1}$ on the right-hand side of equation
(16). One can then use the analogue of equation (14) for 
$\tilde{R}$, $\tilde{R}_{\tau}$ with $c_{k}=\rho_{k}^{D}$ to obtain
the required relations. Expressions for $\rho_{k}^{R}$ and
$\rho_{k}^{R_{\tau}}$ ($k\leq6$) are included in Appendix A.

\section{Numerical evaluation of the contour integral}

In this section we shall reformulate the Taylor's theorem
expansion approach of the last section to obtain a tractable
method for numerically evaluating the contour integral,
appropriate not only when $\tilde{D}$ is truncated at
fixed-order in perturbation theory but crucially also 
allowing us to perform the RS-invariant all-orders
resummation $\tilde{D}^{(L*)}$ of reference \cite {tonge1}.

For ease of reference we shall begin by briefly reviewing
the leading-$b$ resummations. The reader is referred to 
reference \cite {tonge1} for full details.

For a wide range of ``quark-initiated" \cite {benbraun1} 
QCD observables
the perturbative coefficients can be organized as polynomials
in the number of quark flavours, $N_{f}$. That is assuming
such an observable $D(s)$ with the perturbation series
\be
D(s)=a+d_{1}a^{2}+d_{2}a^{3}+\cdots+d_{k}a^{k+1}+\cdots\;, 
\ee 
we can write
\be 
d_{k}=d_{k}^{[k]}N_{f}^{k}+d_{k}^{[k-1]}N_{f}^{k-1}+
\cdots+d_{k}^{[0]}\;.  
\ee 
By substituting $N_{f}$=$(\frac{11}{2}N-3b)$ equation (22)
can be recast in powers of b.
\be
d_{k}=d_{k}^{(k)}b^{k}+d_{k}^{(k-1)}b^{k-1}+\cdots+d_{k}^{(0)}\;.
\ee
Since $d_{k}^{[k]}=(-1/3)^{k}d_{k}^{(k)}$, exact all-orders
large-$N_{f}$ results can be used to perform a ``leading-$b$"
resummation,
\be
D^{(L)}\equiv\sum_{k=0}^{\infty}d_{k}^{(L)}a^{k+1}\;,
\ee
where $d_{k}^{(L)}\equiv d_{k}^{(k)}b^{k}$ ($d_{0}^{(L)}\equiv 1$).

$D^{(L)}$ may be defined as the principal value (PV) 
regulated Borel sum
\be
D^{(L)}(a)=\mbox{PV}\int_{0}^{\infty}\mbox{d}z\,\mbox{e}^{-z/a}
B[D^{(L)}](z)\;,
\ee
$B[D^{(L)}](z)$ denotes the Borel transform, which potentially
involves an infinite set of poles at $z=z_{l}\equiv\frac{2l}{b}$ 
($l=1,2,3,\cdots$) corresponding to infra-red 
renormalons (I${\rm R}_{l}$), and at $z$=$-z_{l}$ corresponding to 
ultra-violet renormalons (U${\rm V}_{l}$). In the specific
case of the Adler $D$-function, $\tilde{D}^{(L)}$, I${\rm R}_{1}$ is
not present reflecting the absence of a relevant operator
of dimension two in the operator product expansion (OPE)
for vacuum polarization \cite {charles1,ben1}. I${\rm R}_{2}$ is a 
single pole
and the remaining singularities are double poles.
Expressions for the residues are given in reference \cite {charles2}.
It is then straightforward to evaluate equation (25) in terms
of the exponential integral functions, 
\be
\mbox{Ei(x)}=-\int_{-x}^{\infty}\mbox{d}t\frac{e^{-t}}{t}\;,
\ee
where for $x>0$ the Cauchy principal value is taken.
The U${\rm V}_{l}$ singularities may then be expressed
in terms of Ei($-Fz_{l}$), where $F\equiv\frac{1}{a}$,
and the  I${\rm R}_{l}$ singularities involve Ei($Fz_{l}$).
Corresponding expressions for $\tilde{D}^{(L)}(F)|_{UV}$
and $\tilde{D}^{(L)}(F)|_{IR}$ are given in equations (48)
and (49) respectively of reference \cite {charles2}.

As pointed out in reference \cite {tonge1} the $D^{(L)}$ 
resummation of 
equation (24) is ambiguous due to its RS-dependence. In 
particular if, as in the case of $\tilde{D}$, the exact NLO
and NNLO coefficients are available it would seem sensible to 
include them and approximate only the unknown $d_{3}, d_{4},
\cdots$, higher coefficients by $d_{3}^{(L)}, d_{4}^{(L)}$.
Unfortunately, however, the resummed result explicitly depends
on the RS chosen for evaluating the exact $d_{1}, d_{2}$
coefficients. This is analogous to the ambiguity encountered in 
matching leading logarithm resummations of jet observables to 
exact fixed-order perturbative results \cite {webber}.
In both cases the
difficulty may be avoided by performing a resummation of the
EC beta-function \cite {tonge1,reader}. The unknown $\rm{N}^{3}$LO and higher
EC beta-function coefficients $\rho_{3},\rho_{4},\cdots$, in
equation (11) are approximated by retaining only the portion
involving the highest power of $b$, 
$\rho_{k}^{(L)}\equiv\rho_{k}^{(k)}b^{k}$. The $\rho_{k}^{(k)}$
can be obtained to all-orders from the large-$N_{f}$ result
for $d_{k}^{(k)}$. If the NNLO invariant $\rho_{2}$ is 
known exactly it can be included. One then arrives at the resummed
EC beta-function 
\be 
\rho^{(L*)}(x)\equiv 
x^{2}(1+cx+\rho_{2}x^{2}+\sum_{k=3}^{\infty}\rho_{k}^{(L)}x^{k})
\;.
\ee

The improved RS-invariant resummation $D^{(L*)}(s)$ can then
be obtained as the solution of the integrated beta-function equation
\be
\frac{1}{D^{(L*)}}+c\ln\frac{cD^{(L*)}}{1+cD^{(L*)}}
\!=\!\frac{b}{2}\ln\frac{s}{\Lambda_{\MS}}-d_{1}^{\MS}(\mu^{2}
\hspace{-1.5mm}=\hspace{-1.5mm}s)-
\int_{0}^{D^{(L*)}}\hspace{-3.5mm}\mbox{d}x
\left[-\frac{1}{\rho^{(L*)}(x)}+
\frac{1}{x^{2}(1+cx)}\right]\;.
\ee
The exact NLO coefficient $d_{1}$ occurs in the RS-invariant
combination \cite {stev1}
\be
\rho_{0}=\frac{b}{2}\ln\frac{s}{\Lambda_{RS}}-d_{1}^{RS}
(\mu^{2}=s)\;.
\ee
The convention assumed for $\Lambda_{\MS}$ in equation (28)
differs from the standard definition by the $N_{f}$-dependent
factor $(2c/b)^{-c/b}$ \cite {stev1}.
If $D^{(L*)}(s)$ is expanded in the coupling $a$ appropriate
to some RS one then obtains
\be
D^{(L*)}\equiv\sum_{k=0}^{\infty}d_{k}^{(L*)}a^{k+1}\,\,\,\,
(d_{0}^{(L*)}=1)\;,
\ee
where now $d_{1}^{(L*)}=d_{1}$ and $d_{2}^{(L*)}=d_{2}$
reproduce the known coefficients and the approximated
$d_{3}^{(L*)}$ and higher coefficients may be obtained in any
RS from the approximated $\rho_{k}^{(L)}$ invariants using
the exact QCD RG. For instance if we label the RS by $d_{1}$
and the beta-function coefficients $c_{2},c_{3},\cdots$,
appearing in equation (13) we have \cite {tonge1} 
\be
d_{3}^{(L*)}(d_{1},c_{2},c_{3})=d_{1}^{3} + 
\frac{5}{2}cd_{1}^{2}+(3\rho_{2}-2c_{2})d_{1}+
\frac{1}{2}(\rho_{3}^{(L)}-c_{3})\;. 
\ee
This differs from the exact $d_{3}$ only in the unknown
$\rho_{3}$ term (we note in passing that $c_{3}^{\MS}$
has now been computed \cite {larin}), the known $\rho_{2}$ has
been exactly included. In this approach the maximum
available exact information is included in an RS-invariant
manner.

It finally remains to perform the resummation $\rho^{(L*)}$
using the explicit expressions for $D^{(L)}(F)$.
Defining
\be 
\rho^{(L)}(x)\equiv 
x^{2}(1+\sum_{k=2}^{\infty}\rho_{k}^{(L)}x^{k})\;,
\ee
and using the chain rule to relate the beta-function
in two different RS's \cite {reader} one has
\be
\rho^{(L)}(x)=\Bigg{(}a^{(L)}(x)\Bigg{)}^{2}\frac{\mbox{d}D^{(L)}(a)}
{\mbox{d}a}\Bigg|_{a=a^{(L)}(x)}\;,
\ee
where $a^{(L)}(x)$ is the inverse function to $D^{(L)}(a)$, i.e. 
$D^{(L)}(a^{(L)}(x))=x$.

$\rho^{(L)}(x)$ can then be straightforwardly defined from 
$D^{(L)}(F)$. The first step is to numerically solve 
\be
D^{(L)}(F(x))=x\;,
\ee
to obtain $F(x)$. Recalling that $F\equiv\frac{1}{a}$ one can
then determine
\be
\rho^{(L)}(x)=-\frac{\mbox{d}}{\mbox{d}F}D^{(L)}(F)
\Bigg|_{F=F(x)}\;,
\ee
by differentiating the explicit $D^{(L)}(F)$ expressions
in terms of $Ei$ functions. Finally on comparing equations
(32) and (27) one has
\be
\rho^{(L*)}(x)=\rho^{(L)}(x)+cx^{3}+\rho_{2}^{(NL)}x^{4}\;,
\ee
where $\rho_{2}^{(NL)}\equiv\rho_{2}-\rho_{2}^{(L)}$.
$\rho^{(L*)}(x)$ can then be 
inserted in equation (28) and the integration performed numerically.
On solving the transcendental equation the RS-invariant 
resummation $D^{(L*)}(s)$ can be evaluated.

We now turn to the problem of evaluating the improved 
resummations $\tilde{R}^{(L**)}$ and 
$\tilde{R}_{\tau}^{(L**)}$ \cite {tonge1} where the 
contour integration
of equation (8) is performed with $D^{(L*)}(se^{i\theta})$
in the integrand,
\be
\hat{R}^{(L**)}(s_{0})=\frac{1}{2\pi}\int_{-\pi}^{\pi}
\mbox{d}\theta\,W(\theta)\,
\tilde{D}^{(L*)}(s_{0}\mbox{e}^{i\theta})\;.
\ee
To perform the contour integration numerically one can
split the range from $\theta=0,\pi$ into $K$ steps of size
$\Delta\theta\equiv\frac{\pi}{K}$,
and perform a sum over the integrand evaluated at 
$\theta_{n}\equiv n\Delta\theta$ $n=0,1,\cdots,K$. So that
\be
\hat{R}(s_{0})\simeq\frac{\Delta\theta}{2\pi}\Big{[}W(0)
\tilde{D}(s_{0})+2\mbox{Re}\sum_{n=1}^{K}W(\theta_{n})
\tilde{D}(s_{n})\Big{]}\;,
\ee
where $s_{n}\equiv s_{0}e^{in\Delta\theta}$.
In practice we perform a Simpson's Rule evaluation.

An efficient strategy \cite {pichcom} is to start with the exact
$\tilde{D}(s_{0})$ and evolve $\tilde{D}(s_{n})$ to 
$\tilde{D}(s_{n+1})$ using Taylor's theorem. Thus
defining $x_{n}\equiv\tilde{D}(s_{n})$ we have
\ba
x_{n+1}&=&x_{n}-\frac{i\Delta\theta}{2}b\rho(x_{n})-
\frac{(\Delta\theta)^{2}}{8}b^{2}\rho(x_{n})
\rho^{\prime}(x_{n}) \nonumber\\
&&\hspace{5mm}+\frac{i(\Delta\theta)^{3}}{48}b^{3}
\Bigg{\{}\rho(x_{n})(\rho^{\prime}(x_{n}))^{2}+(\rho(x_{n}))^{2}
\rho^{\prime\prime}(x_{n})\Bigg{\}}+O(\Delta\theta)^{4}+\cdots\;,
\ea
analogous  to equation (16). If equation (39) is truncated
by retaining its first $m$ terms one anticipates an error
$\sim\frac{1}{K^{m-2}}$ in equation (38).

To evaluate $\tilde{R}_{\tau}$ to four significant figure
accuracy retaining the first four terms in equation (39)
we required 100 steps.

The method obviously can also be used to evaluate the 
contour integral when $\tilde{D}$ is represented by
fixed-order perturbation theory in the coupling
$a(s_{0}e^{i\theta})$,
\be
\tilde{D}(s_{0}e^{i\theta})=a(s_{0}e^{i\theta})+d_{1}a^{2}
(s_{0}e^{i\theta})+d_{2}a^{3}(s_{0}e^{i\theta})\;.
\ee
One can start with $a(s_{0})$ and evolve $a(s_{n})$
to $a(s_{n+1})$ using equation (39) with $\rho(x)$
replaced by the truncated beta-function
in the corresponding RS,
\be
B(x)=x^{2}+cx^{3}+c_{2}x^{4}\;.
\ee
In standard approaches \cite {bert2,racz} the contour 
integral is performed
by solving the integrated beta-function equation with
complex renormalization scale $s_{n}$ for $a(s_{n})$ at each
integration step, and takes much longer to evaluate.
Reference \cite{racz} considers in some detail the RS
dependence of the contour integral.

Determining $\hat{R}^{(L**)}$ with this approach is now
relatively straightforward. For some given $\Lambda_{\MS}$
one evaluates $\tilde{D}^{(L*)}(s_{0})$ \cite {tonge1} as we have
reviewed. The truncated equation (39) is then used
to obtain $x_{1}=\tilde{D}^{(L*)}(s_{1})$. This requires
$\rho^{(L*)}(x_{0})$ and some number of derivatives.
$\rho^{(L)}(x_{0})$ can be determined given the 
$\tilde{D}^{(L)}(F)$ expressions of reference \cite {charles2}, 
and 
using the numerical inversion route of equations (34)
and (35). $\rho^{(L)\,\prime}$, $\rho^{(L)\,\prime\prime}$,
$\cdots$ can then be obtained by successive differentiation
of equation (34) with respect to $x$. One finds
\be 
\rho^{(L)\,\prime}(x)=-\frac{\tilde{D}^{(L)\prime\prime}(F(x))}
{\tilde{D}^{(L)\prime}(F(x))}\;
\ee
and
\be
\rho^{(L)\,\prime\prime}(x)=\frac{\Big{(}
\tilde{D}^{(L)\prime\prime}(F(x))
\Big{)}^{2}}{\Big{(}\tilde{D}^{(L)\prime}(F(x))\Big{)}^{3}}-
\frac{\tilde{D}^{(L)\prime\prime\prime}
(F(x))}{\Big{(}\tilde{D}^{(L)\prime}(F(x))\Big{)}^{2}}\;,
\ee
where primes denote differentiation with respect to $F$.
Thus, once $F(x)$ has been determined from equation (34)
no further transcendental equations need to be solved
and the explicit expressions for $\tilde{D}^{(L)}(F)$
can be repeatedly differentiated to obtain $\rho^{(L)\,\prime}$,
$\rho^{(L)\,\prime\prime}$, $\cdots$. Finally $\rho^{(L*)}$,
$\rho^{(L*)\,\prime}$, $\rho^{(L*)\,\prime\prime}$, $\cdots$,
can be obtained using equation (36) and its derivatives.
For instance
\be
\rho^{(L*)\,\prime}(x)=\rho^{(L)\,\prime}(x)+
3cx^{2}+4\rho_{2}^{(NL)}x^{3}\;.
\ee
The only remaining difficulty is that $x_{1}$ is now
complex, and so at subsequent steps it is unclear
how to obtain $\rho^{(L*)}(x_{n})$, since $\tilde{D}^{(L)}(F)$
is only defined for real $F$. One needs to replace the $\mbox{Ei}(x)$
defined in equation (26) by the generalized exponential integral
functions $\mbox{Ei}(n,w)$ for complex $w$, used to evaluate 
$\tilde{R}^{(L)}$ and $\tilde{R}_{\tau}^{(L)}$ in reference 
\cite {charles2}.
These are defined for $\mbox{Re}\,w>0$ by
\be
\mbox{Ei(n,w)}=\int_{1}^{\infty}\mbox{d}t\frac{e^{-wt}}{t^{n}}\;.
\ee
For $\mbox{Re}\,w<0$ they are defined by analytical continuation
to arrive at a function analytic everywhere in the cut complex
$w$-plane except at $w=0$, and with a branch cut running along
the negative real axis.

To define $\tilde{D}^{(L)}(F)$ correctly for complex $F$
one needs to replace $\mbox{Ei}(-Fz_{l})$ in equation 
(48) of reference
\cite {charles2} for $\tilde{D}^{(L)}(F)|_{\mbox{\tiny{UV}}}$ by 
$-\mbox{Ei}(1,Fz_{l})$. In equation (49) of reference 
\cite {charles2} for 
$\tilde{D}^{(L)}(F)|_{\mbox{\tiny{IR}}}$ one replaces
$\mbox{Ei}(Fz_{l})$ by $-\mbox{Ei}(1,-Fz_{l})+
i\pi\mbox{sign}(\mbox{Im}Fz_{l})$.
In this way as $F$ becomes real one avoids $\pm i\pi$ 
contributions from the discontinuity across the branch cut
along the negative real axis and re-obtains $\tilde{D}^{(L)}(F)$
for real argument.

With $\tilde{D}^{(L)}(F)$ re-defined for complex arguments in 
this way $x_{1},x_{2},\cdots$, can be successively obtained.
At each step one needs to solve the complex-valued 
transcendental equation
\be
\tilde{D}^{(L)}(F_{n})=x_{n}\;,
\ee
and $F_{n}$ is then used to construct $\rho^{(L*)}(x_{n})$
and its derivatives using equations (35), (42) and (43).
The required computing time is dominated by that required for
the solution
of equation (46), and is comparable to that needed
for the conventional approach in fixed-order perturbation
theory, where the complex-valued integrated beta-function
equation is numerically solved at each step.

We have checked that evaluating the contour integral
with $\tilde{D}^{(L)}$ and $\rho^{(L)}$ reproduces
values in numerical agreement with the $\tilde{R}^{(L)}$
and $\tilde{R}_{\tau}^{(L)}$ expressions of reference
\cite {charles2}.

In the next section we shall compare the ``contour-improved"
RS-invariant resummations $\tilde{R}^{(L**)}$ and
$\tilde{R}_{\tau}^{(L**)}$, with ``contour-improved"
fixed-order results obtained by truncating  $\tilde{D}^{(L*)}$
at $n^{th}$ order in the EC scheme, that is in equation (39)
$\rho(x)$ is taken to be the truncation of $\rho^{(L*)}(x)$
in equation (27), retaining terms up to $x^{n}$, and the 
input $\tilde{D}^{(L*)(n)}(s_{0})$ is obtained by solving
equation (28) with the truncated $\rho^{(L*)}(x)$. We shall
denote these by $\tilde{R}^{(L**)[n]}(EC)$ and
$\tilde{R}_{\tau}^{(L**)[n]}(EC)$ for $n\geq3$, and for
$n=1,2$ where the exact $\rho_{k}$ are used by $\tilde{R}^{[n]}(EC)$
and $\tilde{R}_{\tau}^{[n]}(EC)$.

These ``contour-improved" evaluations are to be compared
with conventional fixed-order perturbative truncations
$\tilde{R}^{(L**)(n)}(EC)$ and 
$\tilde{R}_{\tau}^{(L**)(n)}(EC)$ obtained by integrating
up the  $n^{th}$-order truncated EC beta-functions
$\rho^{R,R_{\tau}(L**)}$, with coefficients 
$\rho_{k}^{R,R_{\tau}(L**)}$ obtained using equations (51),
(52) in Appendix A, with the exact $\rho_{2}^{D}$ and 
using $\rho_{k}^{D(L)}$ for $k>2$. By truncating 
$\rho^{R,R_{\tau}}$ one omits an infinite set of 
exactly-known and numerically important analytical
continuation terms which are included in the ``contour-improved"
resummations.

\section{Numerical results}

\setcounter{figure}{0}
\renewcommand{\thefigure}{1(\alph{figure})}

\begin{figure}[t]
\begin{center}
\mbox{\epsfig{file=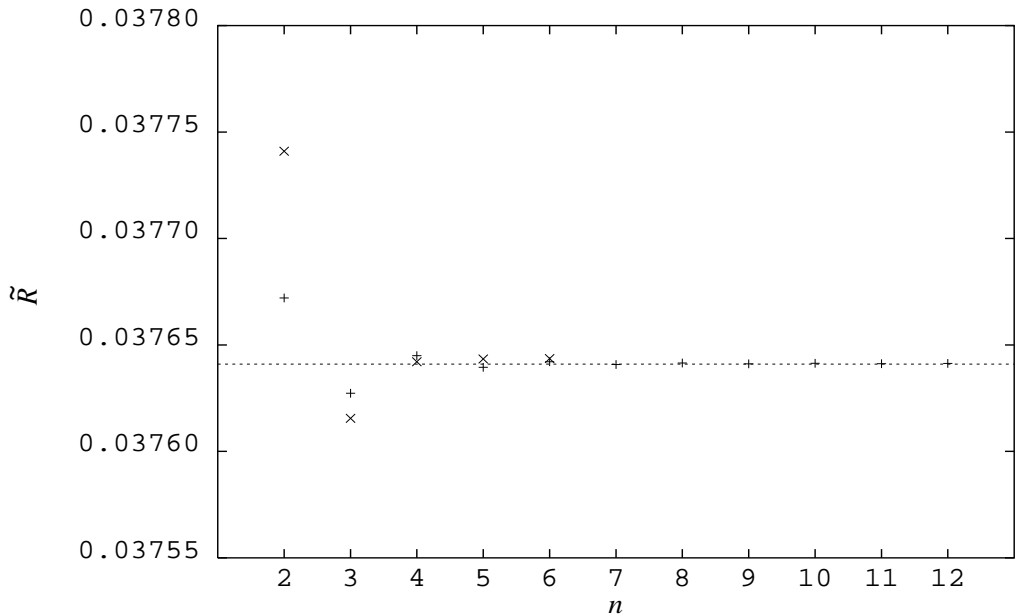,width=5.5in,angle=-90,height=9.3cm}}
\caption{Comparison of two versions of fixed-order EC 
perturbation theory
``contour-improved" $\tilde{R}^{(L**)[n]}(EC)$ (``+") and
$\tilde{R}^{(L**)(n)}(EC)$ (``$\times$") with the 
RS-invariant resummation $\tilde R^{(L**)}(s_{0})$ (dashed line) at
$\surd{s_{0}}=91$ GeV.}
\end{center}
\end{figure}

\begin{figure}[p]
\begin{center}
\mbox{\epsfig{file=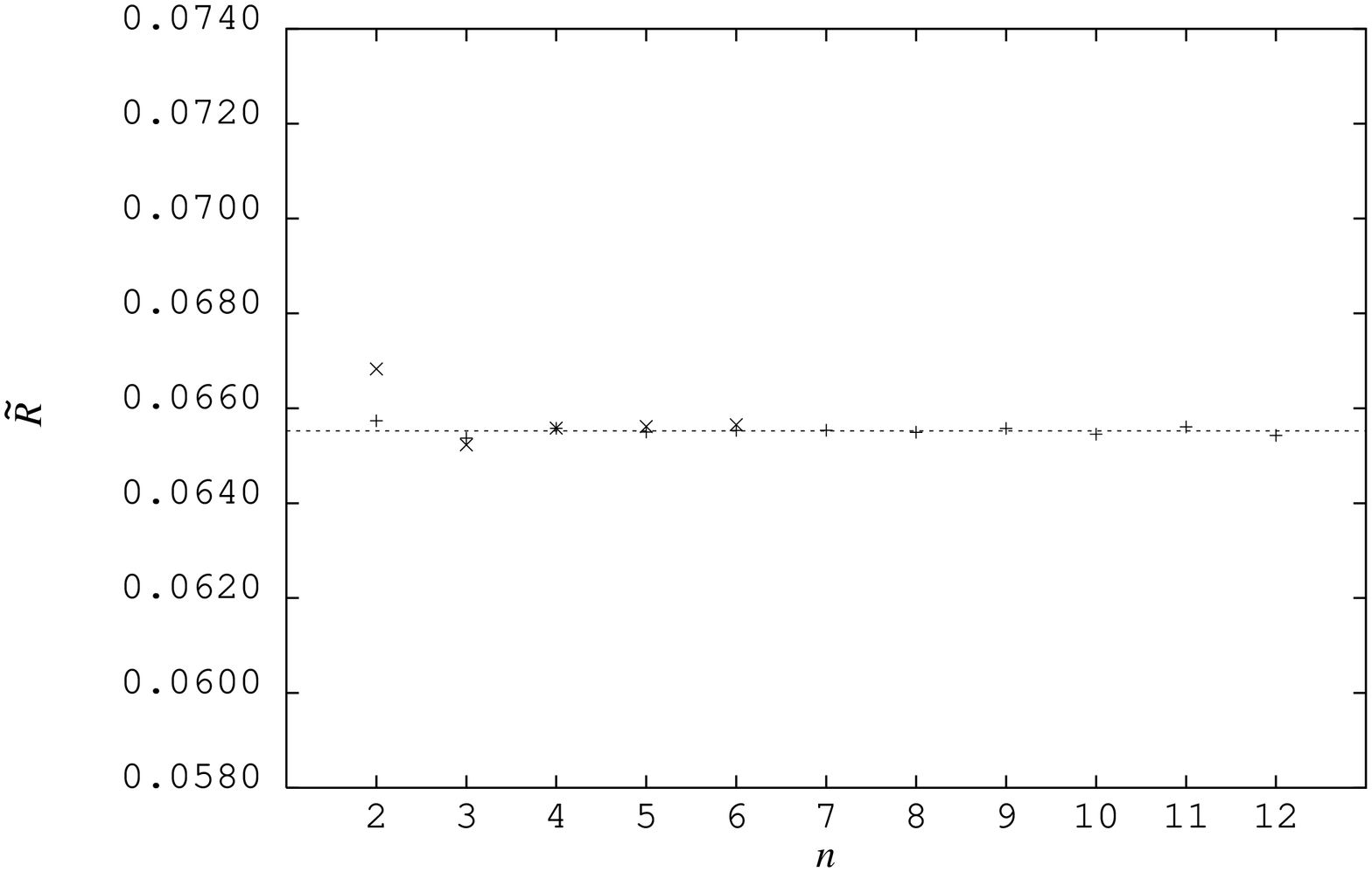,width=5.5in,angle=-90,height=9.3cm}}
\caption{As for Figure 1(a) except at $\surd{s_{0}}=5$ GeV.}
\end{center}
\end{figure}
\begin{figure}[p]
\begin{center}
\mbox{\epsfig{file=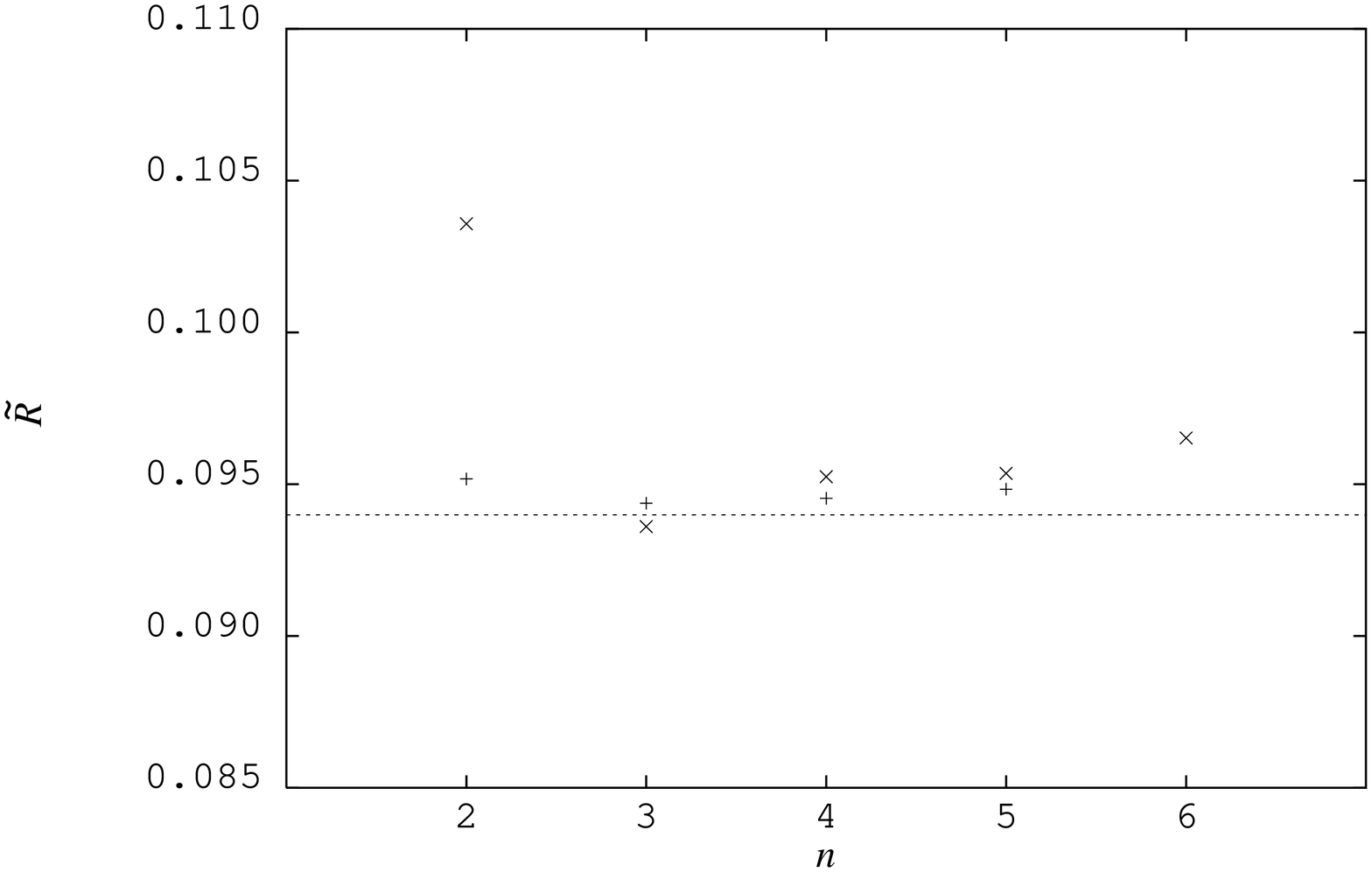,width=5.5in,angle=-90,height=9.3cm}}
\caption{As for Figure 1(a) except at $\surd{s_{0}}=1.5$ GeV}
\end{center}
\end{figure}

In Figures 1(a)-(c), for $\sqrt{s}=91,5,1.5$ GeV, respectively,
we compare the ``contour-improved" resummation 
$\tilde{R}^{(L**)}(s_{0})$ 
with the two versions of fixed-order perturbation theory,
``contour-improved" $\tilde{R}^{(L**)[n]}(EC)$  and
$\tilde{R}^{(L**)(n)}(EC)$ (for $n\geq2$), described in the
last section. Values of $\Lambda_{\MS}^{(5)}=200$ MeV,
$\Lambda_{\MS}^{(4)}=279$ MeV and $\Lambda_{\MS}^{(3)}=320$ MeV
are used. These assume flavour thresholds at $m_{b}=4.5$ GeV
and $m_{c}=1.25$ GeV.

\setcounter{figure}{0}
\renewcommand{\thefigure}{2(\alph{figure})}

\begin{figure}[t]
\begin{center}
\mbox{\epsfig{file=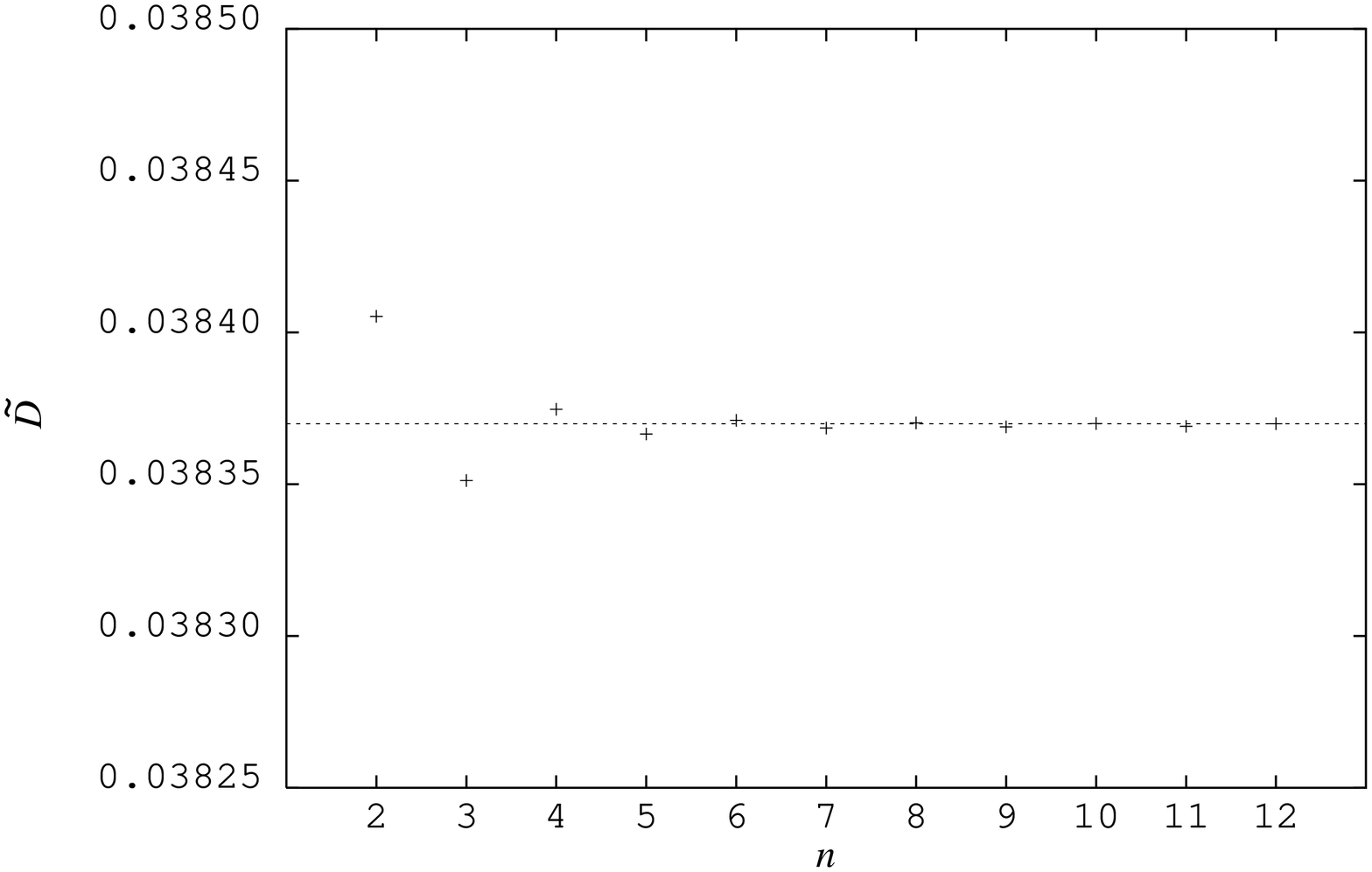,width=5.5in,angle=-90,height=9.3cm}}
\caption{Comparison of fixed-order EC perturbation theory
$\tilde D^{(L*)(n)}(EC)$ (``+") with the RS-invariant resummation
$\tilde D^{(L*)}(s_{0})$ (dashed line) at
$\surd{s_{0}}=91$ GeV.}
\end{center}
\end{figure}
\begin{figure}[p]
\begin{center}
\mbox{\epsfig{file=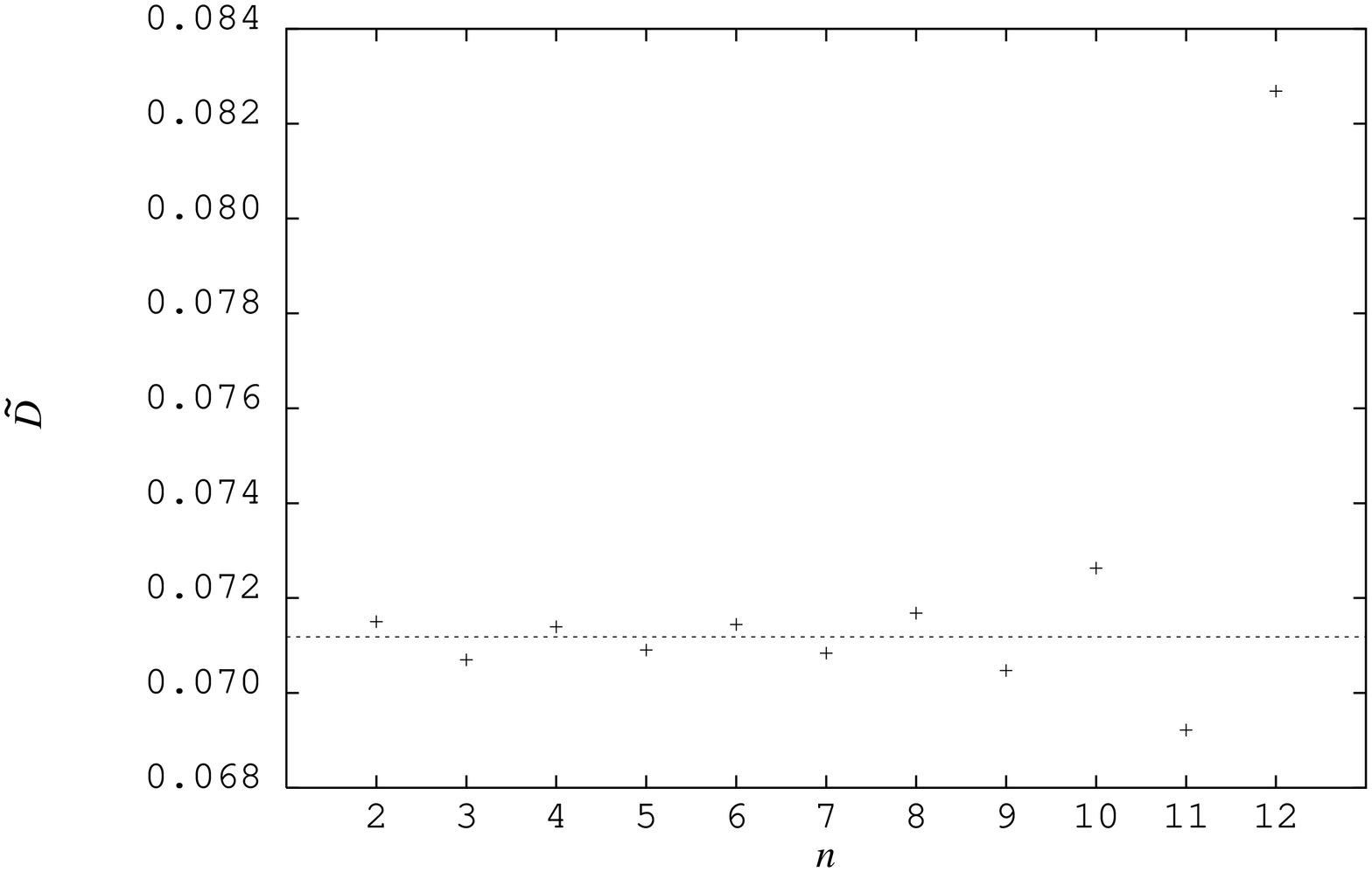,width=5.5in,angle=-90,height=9.3cm}}
\caption{As for Figure 2(a) except at $\surd{s_{0}}=5$ GeV.}
\end{center}
\end{figure}
\begin{figure}[p]
\begin{center}
\mbox{\epsfig{file=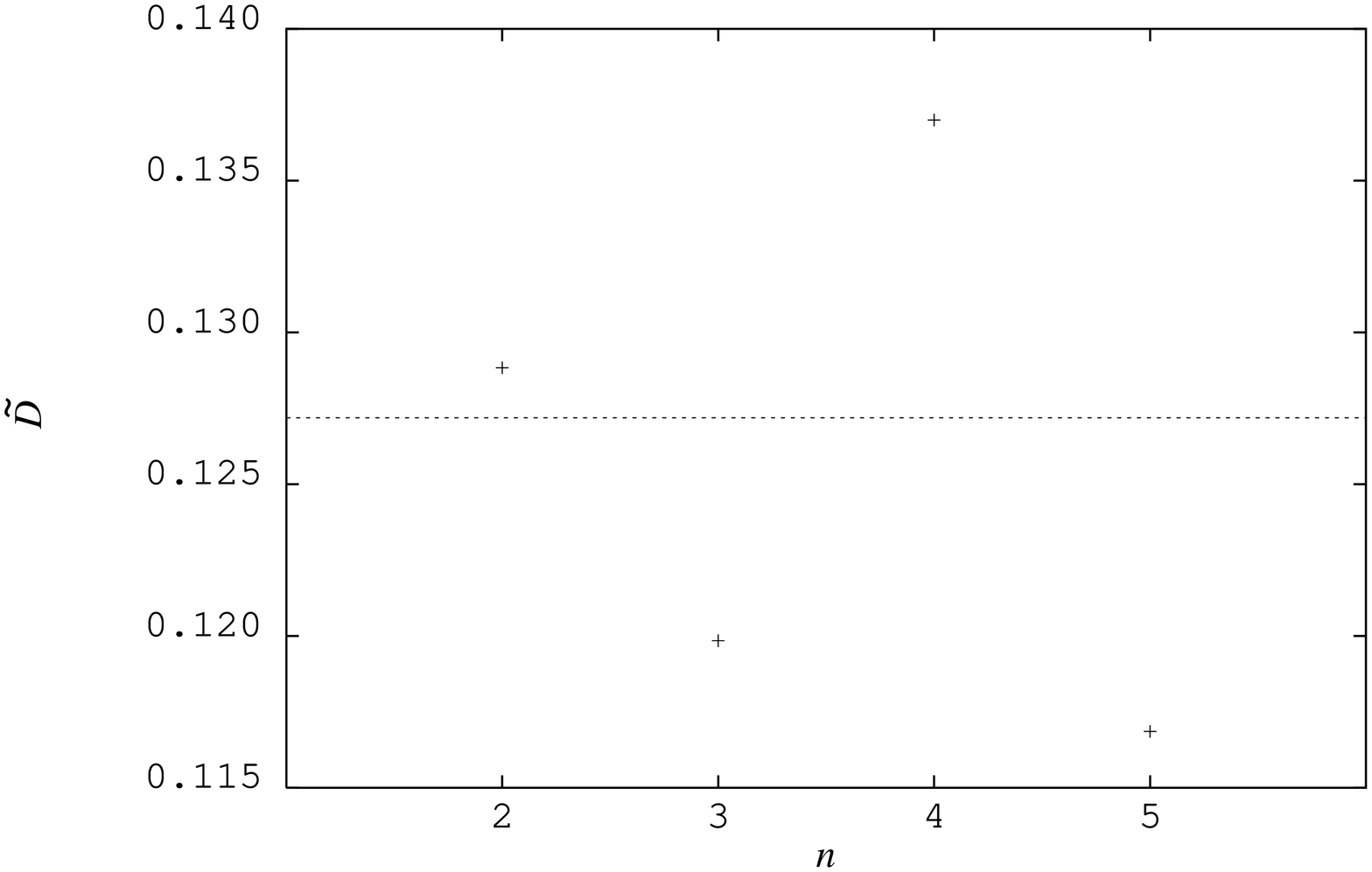,width=5.5in,angle=-90,height=9.3cm}}
\caption{As for Figure 2(a) except at $\surd{s_{0}}=1.5$ GeV.}
\end{center}
\end{figure}

As can be seen at all energies and in low orders the 
``contour-improved" fixed-order results (denoted ``+")
 are significantly closer to the resummation
$\tilde{R}^{(L**)}$ (horizontal line) than the conventional
fixed-order results (denoted ``$\times$''). This is 
completely understandable since both the RS-invariant
resummations and the contour-improved fixed-order results
sum to all-orders known analytical continuation terms, 
as discussed above. The unnecessary truncation of these
terms evidently greatly worsens the performance of $n=2$
NNLO fixed-order perturbation theory, whilst in higher
orders both versions of fixed-order perturbation theory
approach each other, and both track the RS-invariant
resummation. Eventually, of course, both versions will
breakdown as the leading U$\rm{V}_{1}$ renormalon
singularity asserts itself. Since $n=2$ represents
the highest order for which exact calculations exist
at present, ``contour-improvement" is clearly essential
if reliable NNLO determinations of $\alpha_{s}(M_{Z}^{2})$
are to be made.

In Figures 2(a)-(c) we plot $\tilde{D}^{(L*)}(s_{0})$
(dashed line)
and $\tilde{D}^{(L*)(n)}(EC)$ (denoted ``+'') at
$\sqrt{s}=91,5,1.5$ GeV, respectively.
These represent the input values of  $\tilde{D}(s_{0})$
fed into the contour integration to produce the plots in
Figures 1(a)-(c). We note that the fixed-order results in
Figures 2 show a clear oscillation above and below the 
resummed result. This is a reflection of the alternating
sign factorial behaviour contributed by the leading 
U$\rm{V}_{1}$ renormalon, which in the case of $\tilde{D}$
is a double pole, in the leading-$b$ approximation.
A similar oscillatory behaviour is also evident for the 
conventional fixed-order perturbative approximants for
$\tilde{R}$ in Figures 1(a)-(c), but with much smaller
amplitude. This is because for $\tilde{R}$ the U$\rm{V}_{1}$
singularity is softened to a single-pole, again in the 
leading-$b$ approximation. As a result one expects 
$r_{n}/d_{n}\simeq\frac{1}{n}$ asymptotically \cite 
{charles1,ben1},
and correspondingly 
$\rho_{n}^{R}/\rho_{n}^{D}\simeq\frac{1}{n}$. Notice
that the ``contour-improved" fixed-order results which
partially resum higher-order contributions do not exhibit
the simple oscillatory behaviour.

\setcounter{figure}{0}
\renewcommand{\thefigure}{3}

\begin{figure}[t]
\begin{center}
\mbox{\epsfig{file=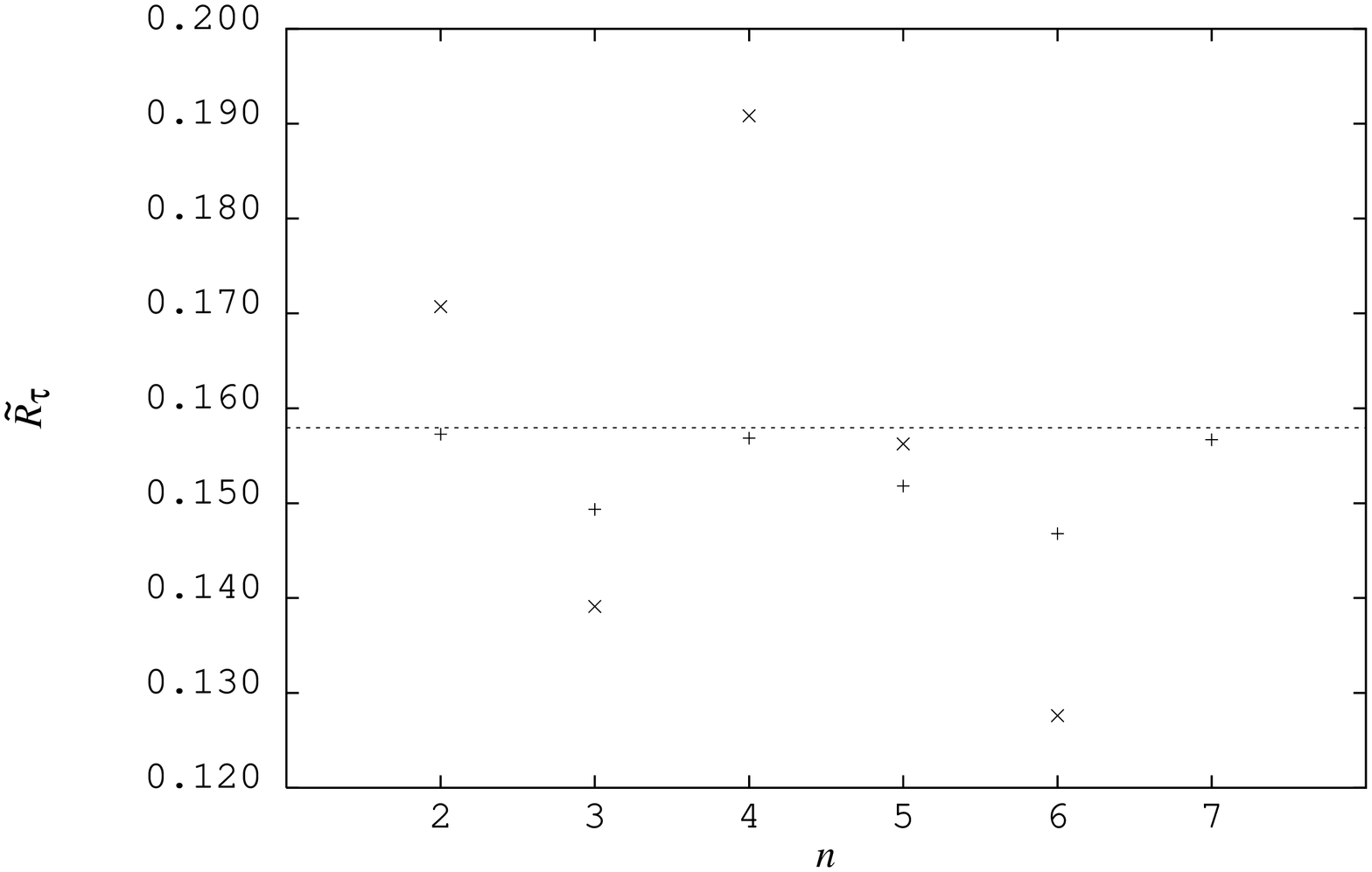,width=5.5in,angle=-90,height=9.3cm}}
\caption{As for Figure 1(a) except for $\tilde{R}_{\tau}$ 
at $\surd{s_{0}}=1.777$ GeV.}
\end{center}
\end{figure}

In Figure 3 we give the analgous plot to Figures 1 for
$\tilde{R}_{\tau}$, assuming $\Lambda_{\MS}^{(3)}=320$ MeV
as before, $\sqrt{s}_{0}=1.777$ GeV. If we compare with
Figure 1(c) for $\tilde{R}$ at the comparable energy
$\sqrt{s}_{0}=1.5$ GeV, we see a deterioration in the 
behaviour of both versions of fixed-order perturbation theory.
The change of weight function from $W(\theta)=1$ to 
$W_{\tau}=(1+2e^{i\theta}-2e^{3i\theta}-e^{4i\theta})$
leads to much less convergent analytical continuation
terms and the two versions of fixed-order perturbation
theory no longer approach each other in higher orders.
The contour-improved results are reasonably close to 
the resummation. Clearly ``contour-improvement" is 
vital for reliable $\alpha_{s}(m_{\tau}^{2})$ determinations.

We now wish to use the difference 
between the 
``contour-improved" 
$\hat{R}^{(L**)}$ and $\hat{R}^{[2]}(EC)$
to estimate the uncertainty with which $\alpha_{s}(M_{Z}^{2})$
can be determined for the Minkowski observables.
Our main interest will be in $\tilde{R}_{\tau}$ which
potentially gives the most accurate determination.
To begin with, however, we consider $\tilde{R}$
at $\sqrt{s}=M_{Z}$ (i.e. the hadronic decay width of
the $Z^{0}$). As in our fits in reference \cite {tonge1} we shall
take $\tilde{R}(M_{Z}^{2})=0.040\pm0.004$. The fits to the
three-loop NNLO $\MS$ $\alpha_{s}(M_{Z}^{2})$ are then
$\alpha_{s}(M_{Z}^{2})=0.122\pm0.012$ from both
$\tilde{R}^{(L**)}$ and $\tilde{R}^{[2]}(EC)$. This is also
the same result as obtained in reference \cite {tonge1} using
$\tilde{R}^{(L*)}$. So at this high energy scale the 
``contour-improvement" has little effect.

For $\tilde{R}_{\tau}$ we take 
$R_{\tau}^{\rm{data}}=3.64\pm0.01$ \cite {bert3}. Correcting for the
small estimated power corrections \cite {bert2} then yields
$\tilde{R}_{\tau}=0.205\pm0.006$. Fitting to the 
``contour-improved" RS-invariant resummation
$\tilde{R}_{\tau}^{(L**)}$ then yields 
$\alpha_{s}(m_{\tau}^{2})=0.339\pm0.006$, and fitting to 
the ``contour-improved" NNLO EC result $\tilde{R}_{\tau}^{[2]}(EC)$
gives $\alpha_{s}(m_{\tau}^{2})=0.350\pm0.008$.
Evolving the three-loop coupling $\alpha_{s}^{\MS}$ using 
Bernreuther-Wetzel matching \cite {BW} from
$N_{f}=3$ to $N_{f}=5$ with the flavour thresholds noted above
yields $\alpha_{s}(M_{Z}^{2})=0.1214\pm0.0007$ and 
$\alpha_{s}(M_{Z}^{2})=0.1226\pm0.0008$, respectively.
Using smaller quark masses ($m_{c}=1.0$ GeV 
and $m_{b}=4.1$ GeV) at the bottom of the range
quoted in \cite {pdg} to perform the evolution yields
$\alpha_{s}(M_{Z}^{2})=0.1222\pm0.0007$ and
$\alpha_{s}(M_{Z}^{2})=0.1234\pm0.0008$, respectively.
Choosing larger masses ($m_{c}=1.6$ GeV 
and $m_{b}=4.5$ GeV) at the top of the quoted range  gives
$\alpha_{s}(M_{Z}^{2})=0.1207\pm0.0007$ and
$\alpha_{s}(M_{Z}^{2})=0.1218\pm0.0008$, respectively.
Thus one can estimate an uncertainty 
$\delta\alpha_{s}(M_{Z}^{2})\simeq0.002$.

In reference \cite {tonge1} we fitted to the same value of 
$\tilde{R}_{\tau}$ using the RS-invariant resummation
$\tilde{R}_{\tau}^{(L*)}$ which only includes
analytical continuation terms at the leading-$b$
level, and found $\alpha_{s}(m_{\tau}^{2})=0.328\pm0.005$,
similarly fitting to NNLO EC fixed-order perturbation theory
gave $\alpha_{s}(m_{\tau}^{2})=0.320\pm0.005$. In both cases
the inclusion of exactly known analytical continuation
terms involving $c, \rho_{2}, \cdots,$ via the
``contour-improvement" serves to significantly increase
the fitted $\alpha_{s}(m_{\tau}^{2})$, and hence 
slightly increase $\alpha_{s}(M_{Z}^{2})$.

\setcounter{figure}{0}
\renewcommand{\thefigure}{4}

\begin{figure}[t]
\begin{center}
\mbox{\epsfig{file=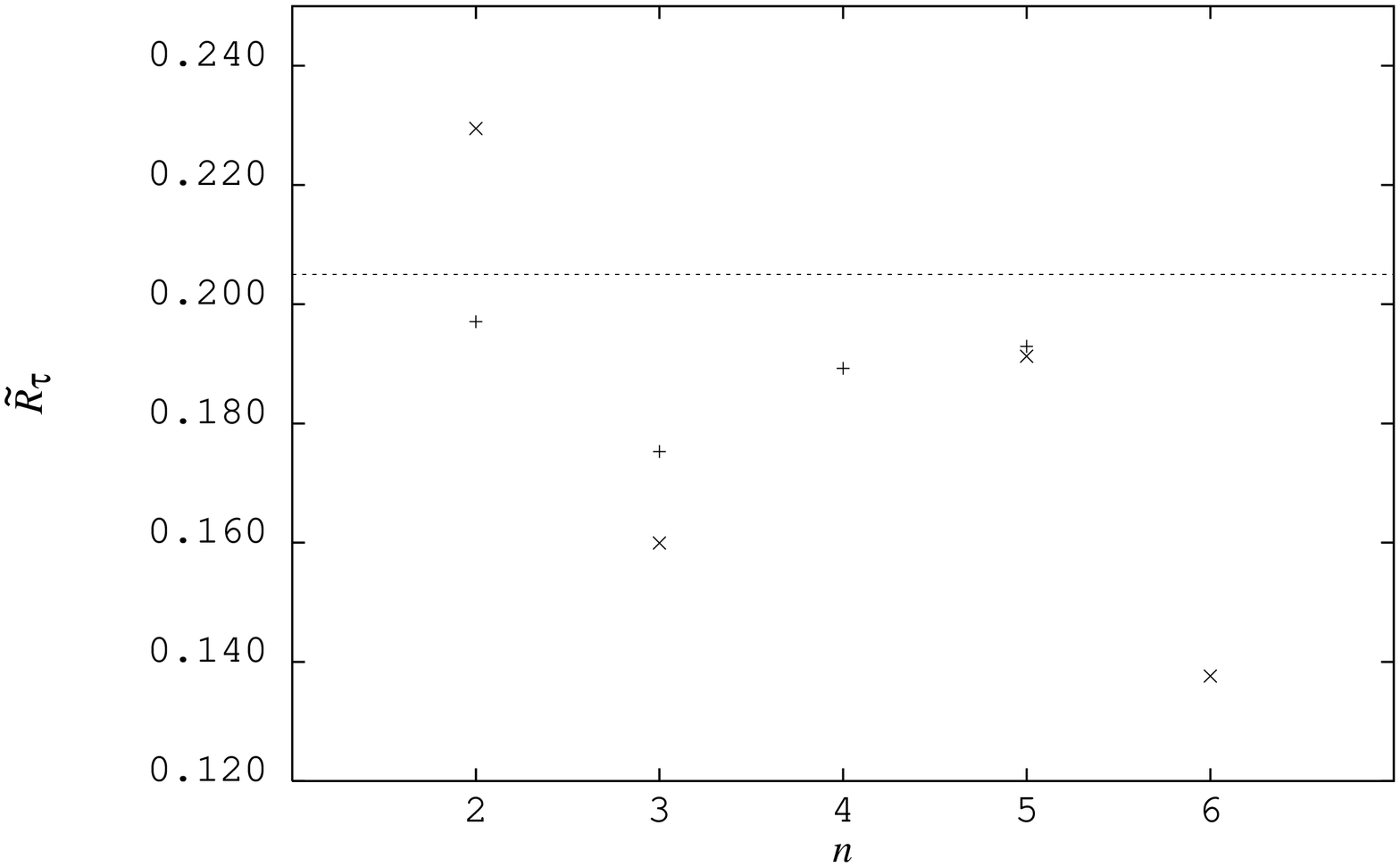,width=5.5in,angle=-90,height=9.3cm}}
\caption{As for Figure 3 except $\tilde{R}_{\tau}^{(L**)}$ 
is fitted to $\tilde{R}_{\tau}^{expt}=0.205$ 
at $\surd{s_{0}}=1.777$ GeV.}
\end{center}
\end{figure}

In Figure 4 we repeat the plot of Figure 3 but with the 
increased value of $\Lambda_{\MS}^{(3)}=429$ MeV, which
results from fitting $\tilde{R}_{\tau}^{(L**)}$ to the data
for $\tilde{R}_{\tau}$. A marked deterioration in the performance
of both versions of fixed-order perturbation theory is evident,
although the ``contour-improved" fixed-order results are still
significantly closer to the resummation. This serves as a 
warning that, at this low energy scale, relatively small 
changes in $m_{\tau}^{2}/(\Lambda_{\MS}^{(3)})^{2}$ can produce
significant changes in the accuracy of perturbation theory.

\setcounter{figure}{0}
\renewcommand{\thefigure}{5}

\begin{figure}[p]
\begin{center}
\mbox{\epsfig{file=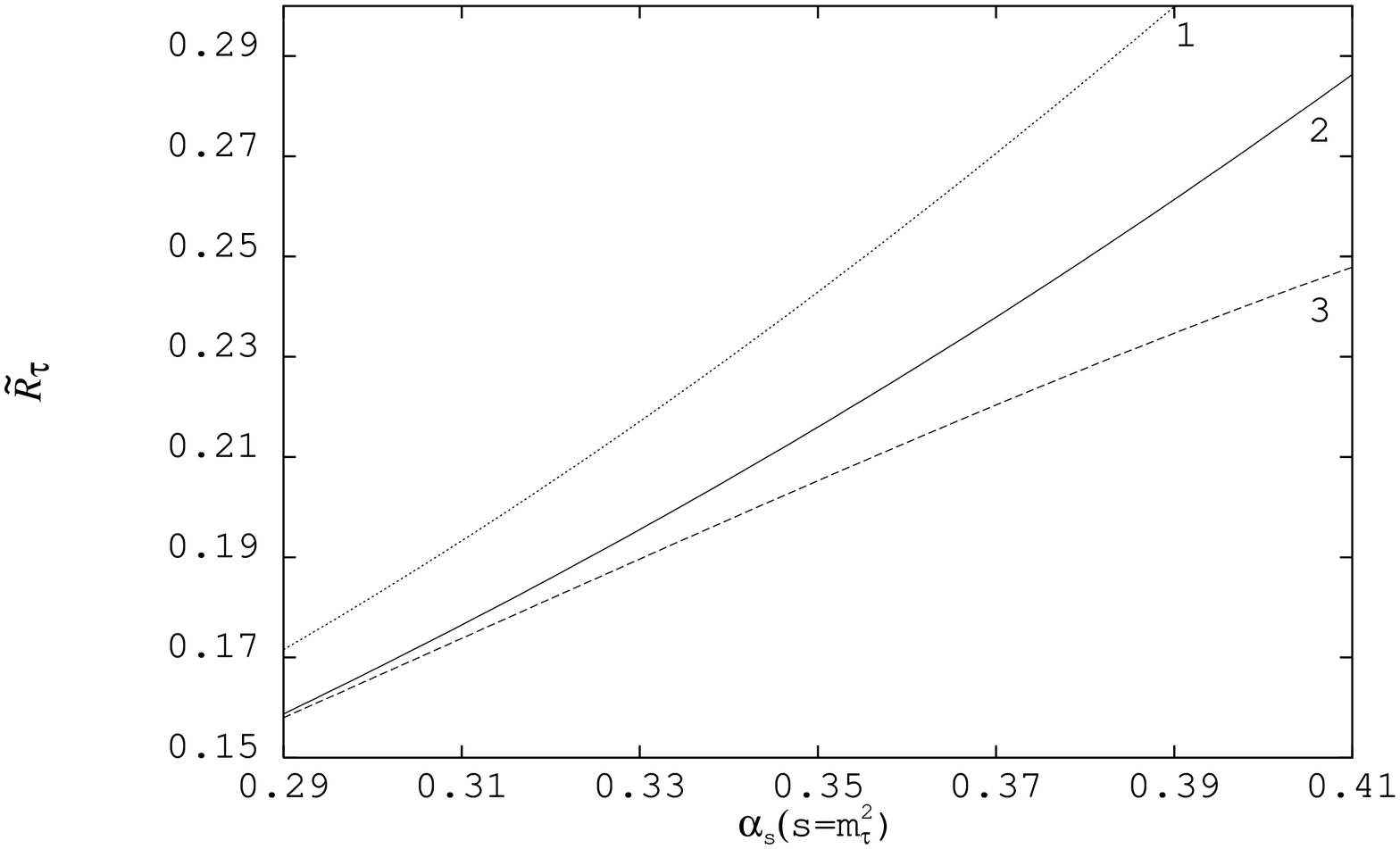,width=5.5in,angle=-90,height=9.0cm}}
\caption{$\tilde{R}_{\tau}$ versus $\alpha_{s}(m_{\tau}^{2})$
for $\tilde{R}_{\tau}^{(2)}(EC)$ (labelled 1),
$\tilde{R}_{\tau}^{(L**)}$ (labelled 2) and 
$\tilde{R}_{\tau}^{[2]}(EC)$ (labelled 3).}
\end{center}
\end{figure}

\setcounter{figure}{0}
\renewcommand{\thefigure}{6}

\begin{figure}[p]
\begin{center}
\mbox{\epsfig{file=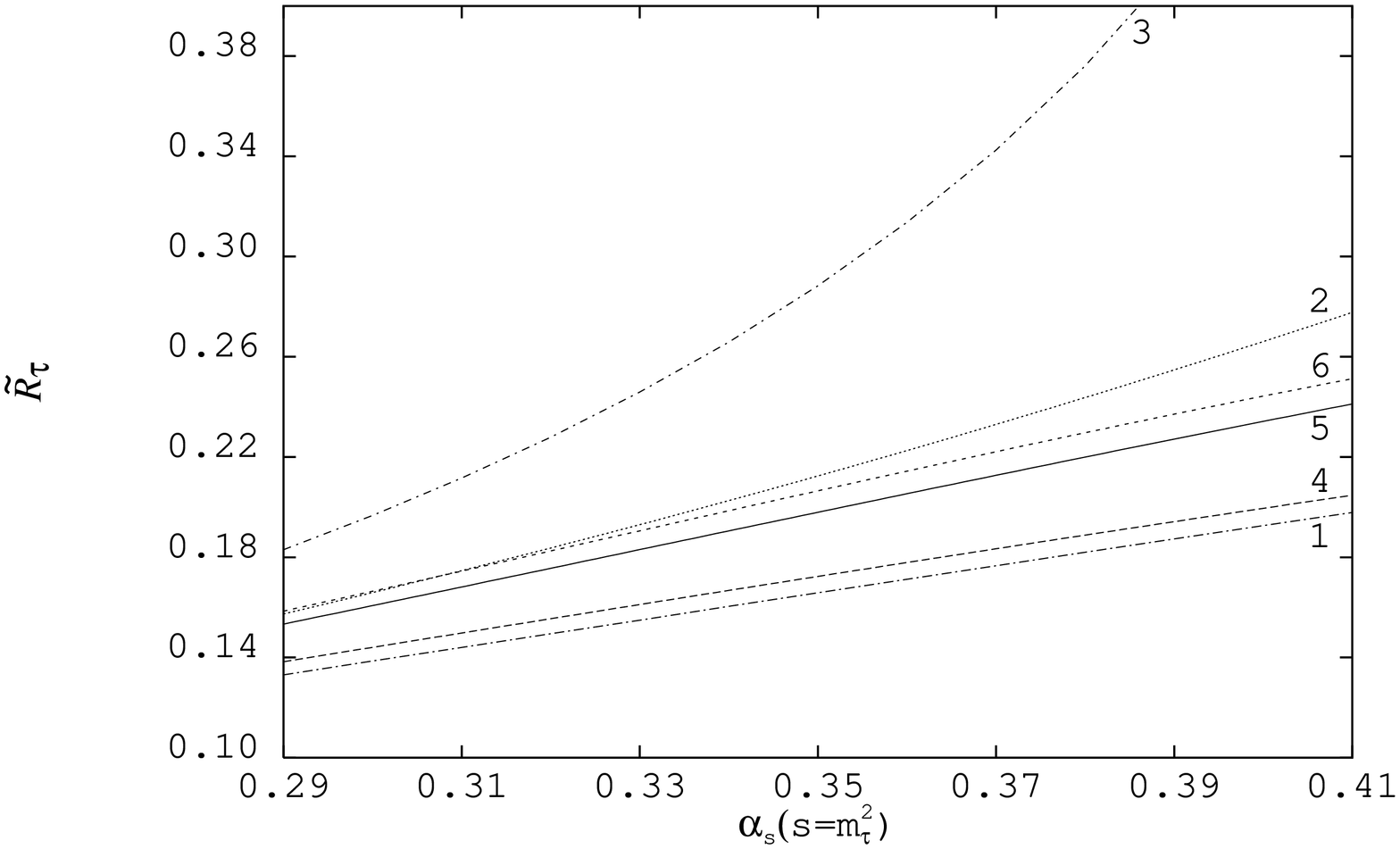,width=5.5in,angle=-90,height=9.0cm}}
\caption[]{$\tilde{R}_{\tau}$ versus $\alpha_{s}(m_{\tau}^{2})$
for $\tilde{R}_{\tau}^{(2)}(\MS)$ and $\tilde{R}_{\tau}^{[2]}(\MS)$,
with three choices of scale $\mu=2m_{\tau}, m_{\tau},
\frac{1}{2}m_{\tau}$; labelled (1, 2, 3) and (4, 5, 6) respectively.}
\end{center}
\end{figure}

In Figure 5 we extend the fits reported earlier. For 
$0.16<\tilde{R}_{\tau}<0.25$ we plot curves for the 
$\alpha_{s}(m_{\tau}^{2})$ obtained by fitting
$\tilde{R}_{\tau}^{(2)}(EC)$ (dotted curve),
$\tilde{R}_{\tau}^{(L**)}$ (solid curve),
$\tilde{R}_{\tau}^{[2]}(EC)$ (dashed curve), to this
value. $\delta\alpha_{s}(m_{\tau}^{2})$ can then be
estimated for given $\tilde{R}_{\tau}$ from the 
difference between the lower two ``contour-improved"
curves. Clearly $\delta\alpha_{s}(m_{\tau}^{2})$
increases very rapidly as $\tilde{R}_{\tau}$
increases. We are very fortunate that apparently
$\tilde{R}_{\tau}\sim0.2$, for which
$\delta\alpha_{s}(m_{\tau}^{2})\simeq0.01$.

We can compare the RS-invariant resummation
$\tilde{R}_{\tau}^{(L**)}$ with the results 
obtained using a Le-Diberder Pich (LP) resummation
\cite {lp}, that is evaluating the contour integral with
$\tilde{D}(m_{\tau}^{2}e^{i\theta})$ as in equation
(40). Fitting $\tilde{R}_{\tau}=0.205\pm0.006$
to this $\tilde{R}_{\tau}^{[2]}(\MS)$ yields
$\alpha_{s}(m_{\tau}^{2})=0.359\pm0.008$
in reasonable accord with the NNLO EC
``contour-improved" value
$\alpha_{s}(m_{\tau}^{2})=0.350\pm0.008$.

Fitting to conventional NNLO fixed-order
perturbation theory in the $\MS$ scheme
$\tilde{R}_{\tau}^{(2)}(\MS)$, with $\mu=m_{\tau}$,
yields $\alpha_{s}(m_{\tau}^{2})=0.342\pm0.006$.
No special significance should be ascribed to the
numerical coincidence that this is close to the 
RS-invariant resummation fit, 
$\alpha_{s}(m_{\tau}^{2})=0.339\pm0.006$.
Crucially these $\MS$ fits are strongly dependent
on the assumed RS. In Figure 6 we show the $\tilde{R}_{\tau}$
versus $\alpha_{s}(m_{\tau}^{2})$ plots with three choices
of scale $\mu=2m_{\tau}, m_{\tau}, \frac{1}{2}m_{\tau}$ 
(labelled 1, 2, 3, respectively). We also plot the curves 
for fitting to a LP resummation based on the NNLO $\MS$ expansion
of $\tilde{D}(m_{\tau}^{2}e^{i\theta})$ in 
$a(4m_{\tau}^{2}e^{i\theta})$, $a(m_{\tau}^{2}e^{i\theta})$
and $a(\frac{1}{4}m_{\tau}^{2}e^{i\theta})$
(labelled 4, 5, 6, respectively). As can be seen the
$\tilde{R}_{\tau}^{(2)}(\MS)$ curves for different scales
are very widely separated. The scale dependence of the LP 
resummations is seen to be much reduced compared to conventional
fixed-order $\MS$ perturbation theory, but is still
significant.

For convenience we now present simple numerical
parametrizations for the contour-improved resummations
$\tilde{R}_{\tau}^{(L**)}$, $\tilde{R}_{\tau}^{[2]}(EC)$
and LP (i.e. $\tilde{R}_{\tau}^{[2]}(\MS)$ based on an
expansion in $a(m_{\tau}^{2}e^{i\theta})$), in terms of
$\alpha_{s}(m_{\tau}^{2})$. We stress that 
$\alpha_{s}(m_{\tau}^{2})$ denotes the 3-loop NNLO $\MS$
coupling with scale $\mu=m_{\tau}$.

Given $x=\tilde{R}_{\tau}$ (data) the fitted 
$\alpha_{s}(m_{\tau}^{2})$ is parametrized by
\be
\alpha_{s}(m_{\tau}^{2})=\pi x+A_{2}x^{2}+
A_{3}x^{3}+A_{4}x^{4}\;.
\ee
The numerical coefficients $A_{i}$ for the different
``contour-improved" versions of perturbation theory
are tabulated in Table 1. These coefficients give
$\alpha_{s}(m_{\tau}^{2})$ to a numerical accuracy
of three significant figures over the range
$0.16\leq\tilde{R}_{\tau}\leq0.25$ covered in Figure 5.

We also present reverse fits. Given 
$x=\alpha_{s}(m_{\tau}^{2})$ the different approximations
for $\tilde{R}_{\tau}$ are parametrized by
\be
\tilde{R}_{\tau}=\frac{1}{\pi}x+\bar{A}_{2}x^{2}+
\bar{A}_{3}x^{3}+\bar{A}_{4}x^{4}\;.
\ee
The numerical coefficients $\bar{A}_{i}$ are 
again tabulated in Table 1. $\tilde{R}_{\tau}$ is accurate
to three significant figures over the range
$0.29<\alpha_{s}(m_{\tau}^{2})<0.41$.

\begin{table}[t]
\begin{center}
\begin{tabular}{|c|c|c|c|c|c|c|}\hline
{\raisebox{-2pt}{Perturbative}} & 
\multicolumn{3}{|c|}{{\raisebox{-2pt}{$A_{i}$}}}&
\multicolumn{3}{|c|}{{\raisebox{-2pt}{$\bar{A}_{i}$}}}\\ 
 \cline{2-7} 
{\raisebox{+1pt}{approximation}} & {\raisebox{-2pt}{$A_{2}$}} 
& {\raisebox{-2pt}{$A_{3}$}}
 & {\raisebox{-2pt}{$A_{4}$}}& {\raisebox{-2pt}{$\bar{A}_{2}$}} &
  {\raisebox{-2pt}{$\bar{A}_{3}$}}
 & {\raisebox{-2pt}{$\bar{A}_{4}$}} \\ 
\hline
{\raisebox{-3pt}{$\tilde{R}_{\tau}^{(L**)}$}} & 
{\raisebox{-3pt}{-13.23}}
 & {\raisebox{-3pt}{+38.87}}
& {\raisebox{-3pt}{-47.44}}& {\raisebox{-3pt}{+.6146}}
 & {\raisebox{-3pt}{+.2099}}
& {\raisebox{-3pt}{+1.352}}  \\
\hline
{\raisebox{-3pt}{$\tilde{R}_{\tau}^{[2]}(EC)$}} & 
{\raisebox{-3pt}{-13.61}} & 
{\raisebox{-3pt}{+39.75}} & {\raisebox{-3pt}{-36.71}} 
& {\raisebox{-3pt}{+.09813}} & 
{\raisebox{-3pt}{+4.504}} & {\raisebox{-3pt}{-7.414}}\\
\hline
{\raisebox{-3pt}{$\tilde{R}_{\tau}^{[2]}(\MS)$}} 
& {\raisebox{-3pt}{-13.92}}
& {\raisebox{-3pt}{+45.55}} & 
{\raisebox{-3pt}{-52.19}} & {\raisebox{-3pt}{+.4319}}
& {\raisebox{-3pt}{+2.121}} & 
{\raisebox{-3pt}{-3.823}}  \\ \hline
\end{tabular}
\caption[t]{Numerical coefficients $A_{i}$ and $\bar{A}_{i}$
 to parametrize $\alpha_{s}(m_{\tau}^{2})$,
 given $x=\tilde{R}_{\tau}$, and $\tilde{R}_{\tau}$, 
 given $x=\alpha_{s}(m_{\tau}^{2})$, respectively, for
 perturbative approximations $\tilde{R}_{\tau}^{(L**)}$, 
 $\tilde{R}_{\tau}^{[2]}(EC)$ and
 $\tilde{R}_{\tau}^{[2]}(\MS)$.}
\end{center}
\end{table}

Finally in this section we wish to examine the 
performance of a straightforward leading-$b$
resummation for $\tilde{R}_{\tau}$. To emphasise
the associated RS ambiguity we shall evaluate
it for three $\MS$ scales $\mu=\lambda m_{\tau}$,
where $\lambda=\frac{1}{2}, 1, 2$, as before.
We then evaluate
\be
\tilde{R}_{\tau}^{(L)}(F(a))+r_{1}^{\tau(NL)}a^{2}+
r_{2}^{\tau(NL)}a^{3}\;.
\ee
Here $a$ denotes the $\MS$ coupling 
$a(\lambda^{2}m_{\tau}^{2})$. $\tilde{R}_{\tau}^{(L)}(F)$
is given by the explicit expressions in equations (69)
and (70) of reference \cite {charles2}. 
$F(a)\equiv\frac{1}{a}-b(\ln{\lambda}+\frac{5}{6})$.
$r_{i}^{\tau(NL)}\equiv r_{i}^{\tau}-r_{i}^{\tau(L)}$.
The extra terms ensure that at NLO and NNLO the known
exact $r_{1}^{\tau}$, $r_{2}^{\tau}$ are included in
the resummation.

Fitting equation (49) to $\tilde{R}_{\tau}=0.205$
as before yields $\alpha_{s}(\frac{1}{4}m_{\tau}^{2})=0.475$,
$\alpha_{s}(m_{\tau}^{2})=0.306$, and
$\alpha_{s}(4m_{\tau}^{2})=0.233$, for the three choices
of RS. Evolving these all to $m_{\tau}$ using the 
three-loop $\MS$ beta-function, which is presumably
appropriate since we are including the exact fixed-order
results to NNLO, one obtains 
$\alpha_{s}(m_{\tau}^{2})=0.297, 0.306, 0.322$, respectively.
Even if we restrict the beta-function to the one-loop
leading-$b$ level, as advocated in references 
\cite {benbraun1,ball,bert2},
we obtain $\alpha_{s}(m_{\tau}^{2})=0.323, 0.306, 0.304$.
Only if we leave out the NL correction terms in equation
(49) and perform a pure leading-$b$ resummation do we
uniquely obtain $\alpha_{s}(m_{\tau}^{2})=0.305$ for all
three RS's.

\section{Discussion and conclusions}

The essential point which motivates our approach
is that the basic ingredient out of which the 
Minkowski observables $\hat{R}$ are built is the
EC beta-function $\rho(\tilde{D}(s))$ defined
in equation (11). Using equation (7) one can see
that this is proportional to 
$\frac{\mbox{d}^{2}}{\mbox{d}\ln{s}^{\,\,2}}\Pi(s)$,
where $\Pi(s)$ is the fundamental correlator of two
vector currents in the Euclidean region defined in 
equation (6). If one specifies $\rho(x)$, then given
the NLO perturbative coefficient $d_{1}^{\MS}(\mu^{2}=s_{0})$
and assuming some value of $\Lambda_{\MS}$, $\tilde{D}(s_{0})$
can be obtained unambigously on solving equation (28).
There is no scale dependence since $\tilde{D}(s_{0})$ only
involves the RS-invariant combination $\rho_{0}(s_{0})$ in
equation (29). Using equation (16) $\hat{R}(s_{0})$
is then also uniquely specified given $\rho(x)$, where
in practice the infinite sum is performed by 
numerically evaluating the contour integral, using
$\rho(x)$ to evolve $\tilde{D}(se^{i\theta})$ around
the circular contour of integration, as described in
section 3.

Of course, the function $\rho(x)$ is not known exactly.
>From NNLO calculations all that is known is the first
three terms in its power series expansion,
\be
\rho(x)=x^{2}+cx^{3}+\rho_{2}x^{4}+\cdots\;.
\ee
The uncertainty in predicting $\hat{R}(s_{0})$ is then
to be estimated by making some approximations for the 
unknown higher order terms indicated by the ellipsis in
equation (50). We have chosen to approximate $\rho_{k}$
by $\rho_{k}^{(L)}$ for $k\geq3$. These leading-$b$
contributions exactly reproduce $\rho_{k}$ in the 
large-$N_{f}$ limit, and for $\rho_{2}$ are a good
approximation in the large-$N$ (or $N_{f}\simeq0$) limit
\cite {tonge1}.
Comparing the predictions for $\tilde{R}_{\tau}$
constructed from the NNLO $\rho(x)$ in equation (50) and
the leading-$b$ resummation indicates a moderate uncertainty
$\delta\alpha_{s}(m_{\tau}^{2})\simeq0.01$ for 
$\tilde{R}_{\tau}\simeq0.2$, which evolves up to 
$\delta\alpha_{s}(M_{Z}^{2})\simeq0.002$ and a central
value $\alpha_{s}(M_{Z}^{2})=0.122$ in line with
other $\alpha_{s}$ measurements, which indicate a
global average $\alpha_{s}(M_{Z}^{2})=0.118\pm0.005$
\cite {burrows}.

Our reassuringly small uncertainty 
$\delta\alpha_{s}(m_{\tau}^{2})\simeq0.01$ is in stark 
contrast to other more pessimistic claims in the
literature. Application of straightforward leading-$b$
resummations compared to exact NNLO fixed-order perturbation
theory leads to a claim of 
$\delta\alpha_{s}(m_{\tau}^{2})\simeq0.05$ in reference \cite {bert3}.
As we showed in section 4, however, there is a matching 
problem if one wishes to include the exactly known
NLO and NNLO coefficients. As a result the 
$\delta\alpha_{s}(m_{\tau}^{2})$ estimate depends 
strongly on the renormalization scale chosen. This
difficulty is avoided in our RS-invariant resummation
approach, and originally motivated it.

In reference \cite {altarelli} an overall uncertainty of 
$\delta\alpha_{s}(m_{\tau}^{2})\simeq0.06$ is claimed.
These authors use an LP resummation together with
an acceleration technique applied to the perturbation
series to lessen the influence of the leading
U$\rm{V}_{1}$ renormalon. The resulting uncertainty
is dominated by the choice of renormalization scale
$\mu$. As we have pointed out above the only uncertainty
in $\tilde{R}_{\tau}$ is due to our lack of knowledge
of the uncalculated RS-invariants 
$\rho_{3}, \rho_{4}, \cdots$. Thus there is no scale 
dependence ambiguity. Since it is $\rho(x)$ which is 
ambiguous one could attempt to improve the convergence
of this series. The corresponding Borel transform has
a U$\rm{V}_{1}$ renormalon and one could try to use
acceleration methods. Crucially, however, the
resulting uncertainties would have to do with real ambiguities
associated with the singularities of the Borel
transform of $\rho(x)$ in the Borel plane, and would not
involve the unphysical and irrelevant renormalization
scale $\mu$. The same criticism applies to reference 
\cite {soper}
which uses similar techniques to assess the perturbative 
ambiguity in $\tilde{R}$.

We therefore conclude that there is no reason to 
suppose that $\tilde{R}_{\tau}$ suffers from serious
ambiguities due to $\rm{N}^{3}$LO and higher terms
which have yet to be exactly calculated. The techniques
on which existing claims to this effect have been based
are all severely RS-dependent, and their conclusions
can be modified at will by making different ad hoc
choices of renormalization scale.

\section*{Acknowledgements}

We would like to thank Andrei Kataev, Toni Pich and
Piotr Raczka for a number of stimulating discussions,
which  have crucially assisted in the development
of these ideas. 

D.G.T gratefully acknowledges receipt of a PPARC 
UK Studentship.

\appendix
\section{Relations between 
RS-invariants for $\tilde R$, $\tilde
R_{\tau}$ and the Adler $D$-function}

Below we present the analytical continuation terms that link
the RS-invariants for the two Minkowski observables
$\tilde R$, $\tilde R_{\tau}$ to those of the Euclidean 
Adler $D$-function.

For $\tilde{R}$ we can relate the Minkowski invariants to the 
Euclidean invariants in the following manner.

\begin{eqnarray}
\rho_{2}^{R}&=&\rho_{2}^{D}-\frac{1}{12}b^{2}\pi^{2}\nonumber\\
\vspace{2mm} \nonumber\\
\rho_{3}^{R}&=&\rho_{3}^{D}-\frac{5}{12}cb^{2}\pi^{2}\nonumber\\
\vspace{2mm} \nonumber\\
\rho_{4}^{R}&=&\rho_{4}^{D}-
\frac{1}{12}(8\rho_{2}^{D}+7c^{2})b^{2}\pi^{2}+
\frac{1}{360}b^{4}\pi^{4} \\
\vspace{2mm} \nonumber\\
\rho_{5}^{R}&=&\rho_{5}^{D}-
\frac{1}{12}(12\rho_{3}^{D}+20\rho_{2}^{D}c+
3c^{3})b^{2}\pi^{2}+\frac{17}{360}cb^{4}\pi^{4}\nonumber\\
\vspace{2mm} \nonumber\\
\rho_{6}^{R}&=&\rho_{6}^{D}-
\frac{1}{12 }(17\rho_{4}^{D}+28\rho_{3}^{D}c+
13(\rho_{2}^{D})^{2}+12\rho_{2}^{D}c)b^{2}\pi^{2}\nonumber\\
& & \;\;\;\;\;+\frac{1}{720}(99\rho_{2}^{D}+137c^{2})b^{4}\pi^{4}-
\frac{1}{20160}b^{6}\pi^{6}\nonumber 
\end{eqnarray}

For $\tilde{R}_{\tau}$ we have more complicated relations.

\begin{eqnarray}
\rho_{2}^{R_{\tau}}&=&\rho_{2}^{D}+I_{2}b^{2}\nonumber\\
\vspace{2mm} \nonumber\\
\rho_{3}^{R_{\tau}}&=&\rho_{3}^{D}+5cI_{2}b^{2}+I_{3}b^{3}\nonumber\\
\vspace{2mm} \nonumber\\
\rho_{4}^{R_{\tau}}&=&\rho_{4}^{D}+(8\rho_{2}^{D}+7c^{2})I_{2}b^{2}+
7cI_{3}b^{3}+I_{4}b^4 \nonumber\\
\vspace{2mm} \nonumber\\
\rho_{5}^{R_{\tau}}&=&\rho_{5}^{D}+(12\rho_{3}^{D}+20\rho_{2}^{D}c+
3c^{3})I_{2}b^{2}+(12\rho_{2}^{D}+16c^{2})I_{3}b^{3} \nonumber\\
& & \;\;\;\;\;+\frac{1}{9}(83I_{4}+28I_{2}^{2})cb^4+I_{5}b^{5} \\
\vspace{2mm} \nonumber\\
\rho_{6}^{R_{\tau}}&=&\rho_{6}^{D}+(17\rho_{4}^{D}+
28\rho_{3}^{D}c\pi^{2}+
13(\rho_{2}^{D})^{2}+12\rho_{2}^{D}c^{2})I_{2}b^{2} \nonumber \\
& & \;\;\;\;\;+\frac{1}{2}(39\rho_{3}^{D}+
99\rho_{2}^{D}c+30c^{3})I_{3}b^{3} \nonumber\\
& & \;\;\;\;\;+\frac{1}{36}(612\rho_{2}^{D}+1081c^{2})I_{4}b^{4}+
\frac{1}{18}(234\rho_{2}^{D}+277c^{2})I_{2}^{2}b^{4} \nonumber\\
& & \;\;\;\;\;+\frac{1}{8}(93I_{5}+60I_{3}I_{2})cb^{5}
+I_{6}b^{6} \nonumber
\end{eqnarray}

where for convenience we have assigned

\begin{eqnarray}
I_{2}&=&\frac{169}{576}-\frac{1}{12}\pi^{2} \nonumber\\
\vspace{2mm} \nonumber\\
I_{3}&=&\frac{1819}{3456} \nonumber \\
\vspace{2mm} \nonumber\\
I_{4}&=&\frac{246779}{165888}-\frac{169}{864}\pi^{2}+
\frac{1}{360}\pi^{4} \nonumber \\
\vspace{2mm} \nonumber\\
I_{5}&=&\frac{269203}{55296}-\frac{1819}{3456}\pi^{2} \nonumber\\
\vspace{2mm} \nonumber\\
I_{6}&=&\frac{392305009}{21233664}-\frac{973531}{442368}\pi^{2}+
\frac{1859}{46080}\pi^{4}-\frac{1}{20160}\pi^{6} \nonumber 
\end{eqnarray}


\newpage
\begin{thebibliography}{99}

\bibitem{lp} F. Le Diberder and A. Pich, 
Phys.Lett. {\bf B289} (1992) 165.
\bibitem{aleph} ALEPH Collaboration, D. Buskulic et al, 
Phys.Lett. {\bf B307} (1993) 209.
\bibitem{cleo} CLEO Collaboration, T. Coan et al, 
Phys.Lett. {\bf B356} (1995) 580.
\bibitem{schilder} K. Schilder and M.D. Tran,
Phys.Rev. {\bf D29} (1984) 570.
\bibitem{braaten1} E. Braaten, 
Phys.Rev.Lett. {\bf 60} (1988) 1606; Phys.Rev. {\bf D39} (1989) 1458.
\bibitem{braaten2} E. Braaten, S. Narison and A. Pich,
Nucl.Phys. {\bf B373} (1992) 581.
\bibitem{piv1} A.A. Pivovarov, Z.Phys. {\bf C53} (1992) 461.
\bibitem{kat} K.G. Chetyrkin, A.L. Kataev and F.V. Tkachov, 
Phys.Lett. {\bf B85} (1979) 277;
M. Dine and J. Sapirstein, Phys.Rev.Lett. {\bf 43} (1979) 668; 
W. Celmaster and R.J. Gonsalves, Phys.Lett. {\bf B44} (1980) 560.
\bibitem{gorish1} S.G.~Gorishny,~A.L.~Kataev~and~S.A.~Larin, 
Phys.Lett. {\bf B259} (1991) 144;\\
L.R. Surguladze and M.A. Samuel, Phys.Rev.Lett. {\bf 66} (1991) 560;
{\bf 66} (1991) 2416 (E).
\bibitem{web} P. Weber, "Review of tau lifetime measurements",
talk given at TAU 96 conference, Colorado, 1996. 
\bibitem{charles1} C.N. Lovett-Turner and C.J. Maxwell, 
Nucl.Phys. {\bf B432} (1994) 147.
\bibitem{charles2} C.N. Lovett-Turner and C.J. Maxwell, 
Nucl.Phys. {\bf B452} (1995) 188.
\bibitem{broad2} D. Broadhurst and A.G. Grozin, 
Phys.Rev. {\bf D52} (1995) 4082.
\bibitem{benbraun1} M. Beneke and V.M. Braun,
Phys.Lett. {\bf B348} (1995) 513.
\bibitem{broad1} D.J. Broadhurst, Z.Phys. {\bf C58} (1993) 339.
\bibitem{ben1} M. Beneke, Nucl.Phys. {\bf B405} (1993) 424.
\bibitem{broadkat} D.J. Broadhurst and A.L. Kataev,
Phys.Lett. {\bf B315} (1993) 179. 
\bibitem{ball} P. Ball, M. Beneke and V.M. Braun,
Nucl.Phys. {\bf B452} (1995) 563.
\bibitem{bert4} M. Neubert and C. Sachrajda,
Nucl.Phys. {\bf B438} (1995) 235.
\bibitem{bert2} M. Neubert, Nucl.Phys. {\bf B463} (1996) 511.
\bibitem{bert3} M. Girone and M. Neubert,
Phys.Rev.Lett. {\bf 76} (1996) 3061.
\bibitem{burrows} P.N. Burrows, Review of $\alpha_{s}$ measurements,
Proc.XXVIII.Int.Conf. on High Energy Physics, Warsaw, July 1996.
SLAC-PUB-7293 (1996).
\bibitem{tonge1} C.J. Maxwell and D.G. Tonge, 
Nucl.Phys. {\bf B481} (1996) 681.
\bibitem{kat2} A.L. Kataev and V.V. Starshenko,
Mod.Phys.Lett. {\bf A10} (1995) 235.
\bibitem{braaten3} E. Braaten and C.S. Li,
Phys.Rev. {\bf D42} (1990) 3888.
\bibitem{sirlin} W.J. Marciano and A. Sirlin,
Phys.Rev.Lett. {\bf 61} (1988) 1815; {\bf 56} (1986) 22.
\bibitem{grun1} G. Grunberg, Phys.Lett. {\bf B95} (1980) 70;
Phys.Rev. {\bf D29} (1984) 2315.
\bibitem{reader} D.T. Barclay, C.J. Maxwell and M.T. Reader,
Phys.Rev {\bf D49} (1994) 3480.
\bibitem{stev1} P.M. Stevenson, Phys.Rev. {\bf D23} (1981) 2916.
\bibitem{piv2} S. Groote, J.G. Korner and A.A. Pivovarov,
[hep-ph/9703268].
\bibitem{racz} P. Raczka and A. Szymacha, Phys.Rev. {\bf D54} 
(1996) 3073;
Z.Phys. {\bf C70} (1996) 125.
\bibitem{webber} S. Catani, G. Turnock and B.R. Webber,
Nucl.Phys. {\bf B407} (1993) 3.
\bibitem{larin} S.A. Larin, T. Van Ritbergen and J.A.M. Vermaseren, 
[hep-ph/9701390].
\bibitem{pichcom} A. Pich, private communication. This approach
is used in the computer program written by F. Le Diberder, widely
applied in experimental analyses.
\bibitem{BW} W. Bernreuther and W. Wetzel,
Nucl.Phys. {\bf B197} (1983) 228.
\bibitem{pdg} Particle Data Group, R.M. Barnett {\it et al.},
Phys.Rev. {\bf D54} (1996) 1.
\bibitem{altarelli} G. Altarelli, P. Narison and G. Ridolfi,
Z.Phys. {\bf C68} (1995) 257.
\bibitem{soper} D.E. Soper and L.R. Surguladze,
Phys.Rev. {\bf D54} (1996) 4566.

\end{thebibliography}
\end{document}